\documentclass[paper]{JHEP3}
\pdfoutput=1

\usepackage{epsfig,bbm}
\usepackage{graphicx}
\usepackage{amsmath}
\usepackage{simplewick}

\newcommand{\be}{\begin{equation}} 
\newcommand{\ee}{\end{equation}}
\newcommand{\bea}{\begin{eqnarray}} 
\newcommand{\eea}{\end{eqnarray}}

\newcommand{\bmp}{\noindent\begin{minipage}{16cm}}
\newcommand{\emp}{\end{minipage}\vskip 7mm} 
\def\lsim{\mathrel{\raise.3ex\hbox{$<$\kern-.75em\lower1ex\hbox{$\sim$}}}}
\def\gsim{\mathrel{\raise.3ex\hbox{$>$\kern-.75em\lower1ex\hbox{$\sim$}}}}

\newcommand{\intron}[1]{}

\newcommand{\mf}{\hat m_f}

\def\ra{{\rightarrow}}

\def\sfrac#1#2{{\textstyle\frac{#1}{#2}}}

\title{Invisible Higgs and Dark Matter}

\author{Matti Heikinheimo\footnote{hmatti@yorku.ca},\\
Department of Physics and Astronomy, York University \\
4700 Keele Street, Toronto, ON, M3J 1P3 Canada}
\author{Kimmo Tuominen\footnote{kimmo.i.tuominen@jyu.fi},
        \\ Department of Physics, P.O.Box 35 (YFL), 
        \\ FI-40014 University of Jyv\"askyl\"a, Finland, 
        \\ and 
  	    \\ Helsinki Institute of Physics, P.O.~Box 64, 
  	    \\ FI-00014 University of Helsinki, Finland.}	    
\author{Jussi Virkaj\"arvi\footnote{virkajarvi@cp3-origins.net},
\\ 
CP$^3$ Origins, Campusvej 55, DK-5230 Odense M, Denmark}

\abstract{
We investigate the possibility that a massive weakly interacting fermion simultaneously provides for a dominant component of the dark matter relic density and an invisible decay width of the Higgs boson at the LHC. As a concrete model realizing such dynamics we consider the minimal walking technicolor, although our results apply more generally. Taking into account the constraints from the electroweak precision measurements and current direct searches for dark matter particles, we find that such scenario is heavily constrained, and large portions of the parameter space are excluded.}


\preprint{CP3-Origins-2012-006 \\ DIAS-2012-7}
\begin{document}
\section{Introduction}

Finding or excluding the Higgs boson of the Standard Model (SM) of elementary
particle interactions is amongst the top priorities of the CERN 
Large Hadron Collider (LHC) experiment. So far the Higgs has escaped direct 
detection, although the year 2011 run culminated in intriguing
hints pointing towards a Higgs boson in the $\sim 125$ GeV mass range 
\cite{ATLAS:2011ae,CMSWW-13Dec2011,Collaboration:2012sm,CMSZZ-13Dec2011}.

In simple extensions of SM, the Higgs can be easily hidden from
the standard searches, explaining the absence of a clear Higgs signal in early 
LHC data. In a typical scenario, the Higgs boson decays dominantly into
a pair of weakly interacting stable particles, which then escape the detector and
are only seen as missing energy. If the production of the Higgs boson is
unaffected by these new physics degrees of freedom, the invisible decays effectively reduce
the cross section of the final states that are looked for in the
standard Higgs searches, since only a subleading part of the total
number of Higgs bosons produced end up in these final states. In this
case the exclusion limits quoted by the experiments do not apply as
such, and a light Higgs boson could still be there. By adjusting the
relative decay width of the invisible channel, some of the excess events
in Higgs boson searches can also be explained \cite{Englert:2011aa}.

The coupling of the Higgs boson contributes to the mass of
the particle it couples to, and hence this new weakly interacting state must be
massive in order to dominate the other decay channels. By construction it is
then a weakly interacting massive particle (WIMP), and could perhaps also provide
plausible dark matter candidate. This scenario has been considered recently
in the case of a scalar singlet dark matter \cite{Raidal:2011xk} and
vector dark matter \cite{Lebedev:2011iq}. 
The case of a singlet fermion was studied in \cite{Baek:2011aa} and \cite{LopezHonorez:2012kv}.

In this paper we point out that this construction may arise as a
consequence of addressing the naturality in the Higgs
sector. A concrete example is provided by  the minimal walking 
technicolor (MWT) model \cite{Sannino:2004qp}
where a fourth generation of leptons is required to cancel the global anomaly
associated with the strongly interacting sector responsible for
electroweak symmetry breaking. 

More generally, we consider an effective theory for a heavy fourth generation lepton doublet 
coupling to an effective SM like (possibly composite) Higgs field.
We investigate if it is possible for the neutrino to provide the dominant decay channel of the
(in this case composite) Higgs boson, and simultaneously produce the right amount of
relic density to contribute the observed dark matter abundance. This
scenario is significantly different from the SM with four sequential generations
(SM4) model, where a complete generation of quarks and leptons is added
to the SM. In the SM4 model the loop induced Higgs coupling to gluons is
enhanced by the two new heavy quarks running in the loop. This effect
significantly enhances the production cross section of the Higgs boson in the
gluon fusion process, and effectively negates the effect that the fourth
generation neutrino could have in hiding the Higgs boson. In the MWT
model there are no fourth generation quarks, so the production cross
section of the Higgs is unchanged, but the decays may be strongly
affected by the new neutrino. For previous results on Higgs decaying into a pair of neutrinos see \cite{Shrock:1982kd}. 
Our main finding is that achieving both the observed 
dark matter relic density and 
strong enough invisible decay width is heavily constrained in this type of model.

In section \ref{sec:model}, we will first review the aspects of the MWT model which are relevant
for the scenario we have outlined in this section. Then, in section \ref{sec:results} we present our numerical results and constraints in light of present data from colliders and dark matter experiments.
In section \ref{sec:checkout} we present our conclusions and outlook. 

%
%

\section{Minimal Walking Technicolor and the Fourth Generation of Leptons}
\label{sec:model}

\subsection{Model Lagrangian and mass terms}

Since we imagine the Higgs to result from a gauge theory confining at the electroweak scale, the low energy degrees of freedom are described better in terms of a chiral effective theory than using the fundamental techniquark and -gluon degrees of freedom. The form of the effective theory is fixed by the underlying chiral symmetry breaking pattern. The simplest possibility is SU(2)$_L\times$SU(2)$_R\rightarrow$SU(2)$_V$, which has three Goldstone bosons. By coupling with the electroweak currents  these are absorbed into the longitudinal degrees of freedom of the weak gauge bosons by the Higgs mechanism, and only the CP-even scalar Higgs remains in the physical spectrum to couple with the matter fermions. 

We denote the left-handed fourth generation lepton doublet by $L_L=(N_L, E_L)$ and the right-handed SU$_L(2)$ singlets as $E_R$ and $N_R$. The interactions between the scalar sector and these leptons up to and including dimension five operators are given by
\begin{eqnarray}
{\mathcal{L}}^{I}_{{\rm{Mass}}} 
  &=& (y \bar{L}_L H E_R+ {\rm{h.c.}})+C_D\bar{L}_L\tilde{H}N_{R}
\nonumber \\
  &+& \frac{C_{L}}{\Lambda}(\bar{L_L}^c\tilde{H})(\tilde{H}^TL_L)
   +  \frac{C_{R}}{\Lambda}(H^\dagger H)\bar{N}^c_{R}N_{R} 
   +{\rm{h.c.}}
\label{scalar_fermion}
\end{eqnarray}
where $\tilde{H}=i\tau^2H^\ast$ and $\Lambda$ is a suppression factor related to the more complete ultraviolet theory like extended Technicolor (ETC) providing a more microscopic origin for these interactions. The first terms in Eq.~(\ref{scalar_fermion}) lead to the usual (Dirac) mass for the charged fourth generation lepton, and the remaining terms allow for more general mass structure of the fourth neutrino. After symmetry breaking the effective Lagrangian (\ref{scalar_fermion}) gives rise to a neutrino mass term:
\begin{equation}  
   -\frac{1}{2}\bar{n}_L^c 
   \left(\begin{array}{cc} M_L & m_D \\ m_D &  M_R\end{array}\right) n_L
   + h.c. \,,
\label{eq:massmatrix}
\end{equation}
where $n_L=(N_{L}, N_{R}^{~c})^T$, $m_D=C_Dv/\sqrt{2}$ and $M_{L,R}=C_{{L,R}}v^2/2\Lambda$, where $v$ is the vacuum expectation value of the effective Higgs field. The special cases are a pure Dirac and a pure Majorana neutrino which are obtained, respectively, by discarding dimension five operators and by removing the right handed field $N_{R}$. The most general mass matrix contains, even after the field redefinitions, one complex phase. However, in this paper we shall restrict ourselves to the case of real mass matrix. Mass eigenstates are two Majorana neutrinos which are related to the gauge eigenstates by a transformation 
\begin{equation}
  N = O n_L + \rho O^T n_L^c \,,
\label{Neigen}
\end{equation}
where $N \equiv (N_1,N_2)^T$ and $O$ is an orthogonal $2\times 2$ rotation matrix, where the associated mixing angle is 
\begin{equation}
  \tan 2\theta = \frac{2m_D}{M_R-M_L} \,.
\label{eq:tan}
\end{equation}
The phase-rotation matrix $\rho = {\rm diag}(\rho_1,\rho_2)$ is included above to ensure that the physical masses $m_{1,2}$ are positive definite. Indeed, the eigenvalues of the mass matrix in (\ref{eq:massmatrix}) are
\begin{equation}
  \lambda_{\pm}=\frac{1}{2}\Big(M_L+M_R\pm\sqrt{(M_L-M_R)^2+4m_D^2} \;\Big) \,.
\label{eq:masseigen}
\end{equation}
Because the signs and relative magnitudes of $M_{L,R}$ and $m_D$ are
arbitrary, the eigenvalues $\lambda_\pm$ can be either positive or
negative. However, choosing  independent phases as
$\rho_{\pm}={\rm{sgn(}}\lambda_{\pm})$ we get positive
$m_\pm=|\lambda_\pm|$ as required. For our purposes it will be
convenient to express everything in terms of the physical mass
eigenvalues $m_1 > m_2$ and the mixing angle $\sin\theta$ instead of the
Lagrangian parameters $M_L$, $M_R$ and $m_D$. While working with
physical parameters has obvious advantages, the downside is that the
connection between the physical and the Lagrangian parameters is not
always straightforward. 

The role of the phase-rotation parameters $\rho_1$ and $\rho_2$ has been
discussed in detail in \cite{Kainulainen:2009rb}. Here we simply
point out the feature that the physically relevant parameter is the
relative sign of $\rho_1$ and $\rho_2$,
\begin{equation}
\rho_1\rho_2\equiv\rho_{12}=\pm 1,
\end{equation}
which divides the parameter space into two parts. The typical limits of
purely left- or right-handed neutrinos and the Dirac-limit are contained
in the $\rho_{12}=-1$ part of the parameter space. We will give our
results for both values of this parameter. 

Naturally also the mixing angle can have positive or negative values depending on the Lagrangian parameters $M_L$, $M_R$ and $m_D$.
However, our results depend only weakly on the sign of the mixing angle and thus we will present our results only for the positive mixing angle values.  

Although the relevant degrees of freedom and their interactions described above are generic and may arise in different beyond the Standard Model scenarios, it is good to have at least one particular microscopic realization available. We consider the MWT model, where the electroweak symmetry breaking is driven by the gauge dynamics of two Dirac fermions in the adjoint representation of SU$_{\rm{TC}}$(2) gauge theory. The key feature of this model is, that it is (quasi) conformal with just one doublet of technifermions \cite{Sannino:2004qp}.  This feature is essential for a technicolor theory not to be at odds with electroweak precision measurements. However, since the technicolor representation is three dimensional, the number of weak doublets is odd and hence anomalous \cite{Witten:fp}. A simple way to cure this anomaly is to introduce one new weak doublet, singlet under technicolor and QCD color \cite{Sannino:2004qp,Dietrich:2005jn} in order not to spoil the walking behavior and to keep the contributions to the oblique corrections as small as possible. Hence, the model requires the existence of a fourth generation of leptons. The anomaly free hypercharge assignments for the new degrees of freedom have been presented in detail in \cite{Dietrich:2005jn}; for the techniquarks we have
\be
Y(Q_L)=\frac{y}{2},\quad Y(U_R,D_R)=\left(\frac{y+1}{2},\frac{y-1}{2}\right),
\ee
while for the fourth generation leptons we have
\be
Y(L_L)=-\frac{3y}{2},\quad Y(N_R,E_R)=\left(\frac{-3y+1}{2},\frac{3y-1}{2}\right).
\ee
In the above equations $y$ is any real number. Choosing $y=1/3$ makes the techniquarks and the new lepton doublet appear 
exactly as a regular standard model family from the weak interactions point of view. The heavy fourth generation neutrino becomes a natural dark matter candidate provided that it is stable; to ensure the stability one can postulate a discrete symmetry.
We note that under this hypercharge assignment fractionally charged techniquark-techniquark or techniquark-technigluon bound 
states may form in the early universe. 
The existence of such charged relics is severely constrained \footnote{Related to this specific bound state of techniquark and 
technigluon there is a recent discussion in \cite{Hapola:2012wi}. }. However, according to Refs.~\cite{Davidson:1991si} and 
\cite{Davidson:2000hf} there seems to exist an open window for relics with masses $\sim$ 0.1-10 TeV. This is exactly the range where 
one would expect the mass of the technihadron states to lay, as the natural scale is of ${\cal{O}}$(TeV). Furthermore, when 
interpreting especially the collider limits, one should keep in mind that, as the relic here is a composite state its production and 
signals in colliders differ from those of charged elementary particles. Additional limits for these relics can follow from WMAP data as 
discussed in \cite{Dubovsky:2003yn}. However, Ref.~\cite{Dubovsky:2003yn}
do not plot their exclusion region to the charge values of our interest. One should also notice, that the WMAP data is always analyzed using specific cosmological model, thus these constraints are model dependent and should not be taken too strictly. Moreover, we do not know for certain if these bounds states would be even created in the first place in the early universe. To give specific constraints for their fate, more complete analysis should be performed e.g. including determination of the interactions of this state with SM particles and the calculation of the relic density. Because this state in the end is part of the other sector (TC) of the underlying model, we postpone this study to future work, and concentrate here to the case where we assume that either this bound state does not exist or that it does not affect the cosmology\footnote{For positive effects of charged relics in cosmology see e.g. \cite{Chuzhoy:2008zy}}.

On the other hand one should notice, that if the hypercharge assignment is changed to $y=1$ there are no fractionally charged 
states, and the fourth generation leptons are
doubly and singly charged and their contribution to the dark matter relic density is disfavored. However, the techniquark-technigluon 
bound state containing $D$-techniquark becomes electrically neutral. This 
state will then be the DM candidate instead of our new neutrino. 
This kind of model has been studied previously in \cite{Kouvaris:2007iq}. In this case our analysis and results presented here can be 
directly interpret as an analysis for the techniquark-technigluon dark matter. A somewhat similar scenario of a composite DM particle from a strongly interacting hidden sector has been considered in \cite{Hur:2011sv}. As yet another model building alternative, instead of 
single lepton doublet with hypercharge $Y(L_L)=-3/2$,  one can saturate the Witten anomaly by introducing three SM-like lepton 
doublets, and a stable heavy neutrino among these three becomes a plausible dark matter candidate. Various further possibilities of 
nonsequential generations beyond the SM have been considered in literature \cite{Antipin:2010it,Knochel:2011ng}.

Finally we note that in MWT, the global symmetry breaking pattern is SU$(4)\rightarrow {\rm SO}(4)$, with nine Goldstone bosons. 
Three of these are absorbed into the longitudinal degrees of freedom of the weak gauge bosons, and the low energy spectrum is 
expected to contain six quasi Goldstone bosons which receive mass through extended technicolor interactions
~\cite{Hill:2002ap,Appelquist:2002me,Appelquist:2004ai}. Their phenomenology has been investigated elsewhere 
\cite{Foadi:2007ue,Foadi:2008ci,Hapola:2012wi}. For the hypercharge assignements we consider,  the conservation of hypercharge allows only the effective 
SM-like scalar Higgs to couple to the fermions. Hence, to consider the interactions between fermions and the scalar sector, the 
interactions introduced in Eq. (\ref{scalar_fermion}) are sufficient.

Of course it is possible that the (composite) Higgs is not the only source of the fermion masses, but  there are other (composite or even fundamental) scalars whose condensation leads to mass terms for the matter fields. To illustrate such possibilities, we consider as an alternative to the model Lagrangian~(\ref{scalar_fermion}), the case where the right-handed neutrino mass originates from a Standard Model singlet scalar field $S$. 
\begin{eqnarray}
{\mathcal{L}}^{II}_{{\rm{Mass}}} 
  &=& (y \bar{L}_L H E_R+ {\rm{h.c.}})+C_D\bar{L}_L\tilde{H}N_{R}
\nonumber \\
  &+& \frac{C_{L}}{\Lambda}(\bar{L_L}^c\tilde{H})(\tilde{H}^TL_L)
   +  C_{R}S\bar{N}^c_{R}N_{R} 
   +{\rm{h.c.}}
\label{scalar_fermion2}
\end{eqnarray}
This model is similar to the usual see-saw neutrino mass generation mechanism, although here the singlet $S$ does not need to be a fundamental scalar. To specify the model completely one should give a potential for $S$. However, none of the parameters of this potential are needed in our analysis since we may assume that the vacuum expectation value for $S$ is generated through interactions with the Higgs, i.e. we do not need additional sources of spontaneous symmetry breaking for the dynamics of the $S$-field. In what follows, we will refer to the scenario with just only the doublet Higgs field as Scenario I and to the case with Higgs and a singlet scalar as Scenario II.

\subsection{Couplings}
\label{sec:couplings}

For the analysis of the Higgs decay branching ratios and relic density
we need the couplings of the neutrino mass eigenstates to the Higgs
boson and to the weak gauge bosons. These are easily found out by applying the appropriate phase- and rotation transformations defined in the previous section. We shall write down only the terms relevant for our calculations.  For the $Z$ and $W^\pm$ bosons we find that
\begin{eqnarray}
W^+_\mu\bar{N}_L\gamma^\mu E_L 
  &=& \sin\theta \; W^+_\mu \bar{N}_{2L}\gamma^\mu E_L+\cdots  \nonumber\\
Z_\mu\bar{N}_L\gamma^\mu N_L 
  &=& \sin^2\theta \; Z_\mu \bar{N}_{2L}\gamma^\mu N_{2L} \nonumber \\
  &+& \sfrac{1}{2}\sin 2\theta \;  Z_\mu \, ( \bar{N}_{1L}\gamma^\mu N_{2L}                              
                             + \bar{N}_{2L}\gamma^\mu N_{1L} ) + \cdots  \,, 
\label{eq:intlagrange}
\end{eqnarray}
where the omitted terms contain interactions of the heavy $N_1$ field only. These couplings are diagonal in the mixing and therefore do not involve the phase factor $\rho_{12}$. However, neutral current involves mixing and these couplings do depend on $\rho_{12}$. One finds:
\begin{equation}
\bar{N}_2\gamma^\mu Z_\mu P_L N_1+\bar{N}_1\gamma^\mu Z_\mu P_LN_2 = \bar{N}_2(\beta+\alpha\gamma_5)\gamma^\mu Z_\mu N_1,
\label{eq:neutralcurrent}
\end{equation}
where 
\begin{equation}
\alpha=\sfrac{1}{2}(1+\rho_{12}) \quad {\rm  and} \quad \beta=\sfrac{1}{2}(1-\rho_{12}) \,.
\label{eq:alphabeta}
\end{equation}
Thus, for $\rho_{12} = -1$ the neutral current interaction of our WIMP is purely axial vector and for $\rho_{12} = +1$ purely vector. Usually in the literature dealing with the interactions of Majorana neutrinos, only the first possibility is mentioned, although e.g. \cite{LopezHonorez:2012kv} considers both operators.

The effective interaction terms involving the Higgs and the lighter neutrino eigenstate are
\begin{eqnarray}
{\mathcal{L}}_{NH}  &=& 
\frac{gm_2}{2M_W}\Big( \; C_{22}^h h\bar{N_2}N_2+C_{21}^h h\bar{N_1}(\alpha-\beta\gamma^5)N_2 
\nonumber\\ 
&& \phantom{Ham} + \frac{1}{v}C_{22}^{h^2} h^2\bar{N_2}N_2 \,\Big)+
\frac{m_H^2}{2v}h^3 + \cdots \,,
\label{higgs_interactions}
\end{eqnarray}
where we have again omitted the interaction terms which do not contain
$N_2$ and hence are not needed in our analysis. 
%
%
%
%

The interactions between the Higgs and the neutrino can be generically described by the Lagrangian (\ref{higgs_interactions}) for scenarios I and II we have introduced. The factors $\alpha$ and
$\beta$ are  defined in Eq. (\ref{eq:alphabeta}) and the factors
$C^h_{22}$, $C^h_{21}$ and $C^{h^2}_{22}$ are given in
Table~\ref{c_table}. 
\TABLE[t]{
\begin{tabular}{| l | c | c |}
\hline
 & Scenario I & Scenario II \\
\hline
$C_{22}^h$     & $1-\frac{1}{4}\sin^22\theta\, R_{-}$ 
               & $\sin^2\theta$ \\ [2mm]
$C_{21}^h$     & $-\frac{1}{4}\rho_{12}\sin4\theta\,R_-$ 
               & $ \frac{1}{2}\rho_{12}\sin2\theta\,R_+$    \\[2mm]
$C_{22}^{h^2}$ & $\frac{1}{2}-\frac{1}{4}\sin^22\theta\, R_{-}$ 
               & $\frac{1}{2}\sin^2\theta (1-\cos^2\theta R_-) $ \\ [1mm]
\hline
\end{tabular}
\caption{ Coefficients of the Lagrangian (\ref{higgs_interactions}) for two the distinct mass generating scenarios described by Eqs.~(\ref{scalar_fermion}) and (\ref{scalar_fermion2}). We have defined $R_{\pm}\equiv 1 \pm \rho_{12}\frac{m_1}{m_2}$. }
\label{c_table}
}

\subsection{Oblique constraints}
\label{sec:oblique}

The fourth generation of leptons is constrained by current accelerator
data. From LEP we know that the charged lepton $E$ has to be more
massive than the $Z$ boson and if the fourth generation neutrino has
standard model interaction strength, it needs to be heavier than $M_Z/2$
in order to evade the constraint from $Z$-pole observables. In the case
of neutrino mixing considered in this work, the lighter state can have a
substantial right-handed component and hence interact only very
weakly. This could allow this state to escape the LEP bounds even when
its mass is less than $M_Z/2$ (see e.g. \cite{Kainulainen:2009rb}), but in this work we will limit to the
case $m_2>M_Z/2$. In addition to these direct bounds, the parameters of the fourth generation leptons are constrained by oblique corrections, i.e. due to their contribution to the vacuum polarizations of the electroweak gauge bosons. These contributions are conveniently represented by the $S$ and $T$ parameters \cite{Peskin:1990zt}.   

The oblique corrections in MWT model with the general mass and mixing
patterns considered here have been studied in detail in
\cite{Antipin:2009ks} and \cite{Frandsen:2009fs}. We also note that there exists two extensive fits
performed by the LEP Electroweak Working Group (LEPEWWG) \cite{:2005ema}
and independently by the PDG \cite{Nakamura:2010zzi}. Both fits find that the SM, defined to lie at
$(S,T)=(0,0)$ with $m_t=170.9$ GeV and $m_H=117$ GeV, is within
$1\sigma$ of the central value of the fit. The two fits disagree slightly on the
central best-fit value: LEPEWWG finds a central value
$(S,T)=(0.04,0.08)$ while including the low energy data the PDG  finds
$(S,T)=(0.03,0.07)$.  Since the actual level of coincidence inferred
from these fits depends on the precise nature of the fit, we allow a
broader range of $S$ and $T$ values, roughly corresponding to the
$3\sigma$ contour. From the results of \cite{Antipin:2009ks} it can be
inferred that these values can be accommodated easily within the
parameter space of the leptonic sector.

\begin{figure}
\begin{center}
\includegraphics[width=0.75\textwidth]{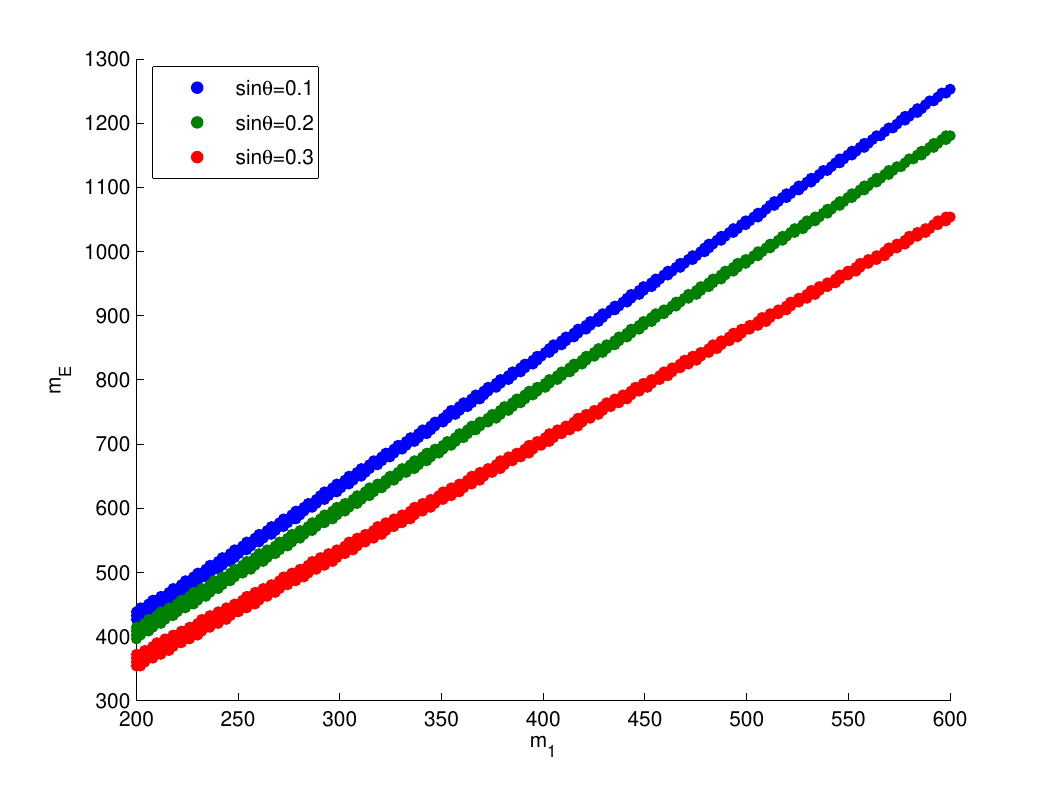}
\caption{The allowed mass spectrum of the fourth generation leptons for
 $m_2=62$ GeV, $\rho_{12}=-1$ and $\sin\theta=0.1, 0,2$ or $0.3$.}
\label{fig:masses}
\end{center}
\end{figure}

For reference we show the allowed mass spectrum of the fourth generation
in figure \ref{fig:masses} for a choise of the lightest neutrino mass
$m_2=62$ GeV, and $\sin\theta=0.1, 0.2$ or $0.3$. As was discussed in
\cite{Antipin:2009ks}, the limiting factor is the $T$-parameter, which
fixes the ratio of the two masses $m_1$ and $m_E$ to a narrow
range. This ratio is slightly dependent on the value of the mixing
angle, as can be seen in figure \ref{fig:masses}. Here we have used $\rho_{12}=-1$. 
For $\rho_{12}=+1$ the $T$-parameter is generally a bit smaller and thus the allowed range of masses is slightly wider. The absolute scale of
the masses affects the $S$ parameter slightly, but the constraint from
$S$ is much weaker than the one from $T$.

Here we are interested in finding parts of the parameter space, where
the precision constraints are met, the lightest neutrino mass
eigenstate provides the correct relic density to match the observed DM
abundance, and the invisible decay channel $H\rightarrow N_2N_2$ is the
dominant decay channel of the composite Higgs boson. Our strategy for
scanning the parameter space is as follows: We scan the two-dimensional
parameter space defined by the mass of the lightest neutrino mass
eigenstate $m_2$ and the neutrino mixing angle $\sin\theta$. For each
point in this plane we scan over the subspace defined by the mass
of the heavier neutrino $m_1$ and the mass of the charged lepton $m_E$,
and select the point in this plane that gives the most suitable value
for the oblique parameters $S$ and $T$. In the case $\rho_{12}=+1$,
practically the whole $(m_2,\sin\theta)$-plane is allowed in terms of
the oblique constraints. That is, for every point in that plane there is
a configuration of the values of $m_1$ and $m_E$ that produces
acceptable values for $S$ and $T$. If $\rho_{12}=-1$ the $T$ parameter
gets larger with large values of $\sin\theta$ and the values above
$\sin\theta\gtrsim 0.45$ are ruled out.

We then calculate the invisible decay width of the composite Higgs boson
and the relic density of the light neutrino in this point of the
parameter space. In the following sections we will describe the
evaluation of the relic density, the constraints from earth-based direct
dark matter searches, and the calculation of the invisible decay width
of the Higgs boson. We will then present the results in section \ref{sec:results}.

\subsection{Higgs Decay}

The Higgs coupling to the lightest neutrino is given in equation
(\ref{higgs_interactions}). The resulting tree-level decay width is
\begin{equation}
\Gamma_{H\rightarrow N_2N_2}=\frac{G_F(m_2C^h_{22})^2 m_H}{2\pi\sqrt2}\left(1-\left(\frac{2m_2}{m_H}\right)^2\right)^\frac{3}{2}.
\end{equation}
The invisible branching ratio of the composite Higgs boson is defined as
\begin{equation}
R_\Gamma=\frac{\Gamma_{H\rightarrow N_2N_2}}{\Gamma_{H\rightarrow
 N_2N_2}+\Gamma_{H}^{\rm SM}},
\end{equation}
where $\Gamma_{H}^{\rm SM}$ is the total Higgs decay width in the
SM. The relative decay width to a given SM decay channel $H\rightarrow XX$ is then
modified by a factor of $(1-R_\Gamma)$:
\begin{equation}
R_{XX}=\frac{\Gamma_{H \rightarrow XX}}{\Gamma_{H\rightarrow N_2N_2}+\Gamma_{H}^{\rm
 SM}}=(1-R_\Gamma)R_{XX}^{\rm SM},
\end{equation}
where $R_{XX}^{\rm SM}=\Gamma_{H \rightarrow XX} / \Gamma_{H}^{\rm SM}$ is the
corresponding branching ratio in the SM. Since the production cross
section of the Higgs boson in our model is equal to the SM, this
suppression effectively results in a suppression of the total cross
section for a given Higgs boson search channel.

\subsection{Relic Density}
\label{sec:omega}

Here we will summarize and update the relic density analysis which is originally given for this particular model in work \cite{Kainulainen:2009rb}.
We calculate the DM abundance $\Omega_{N_2}$ in the standard way, using the Lee-Weinberg equation~\cite{Lee:1977ua} for scaled WIMP number density:
\begin{equation}
\frac{\partial f(x)}{\partial x} 
        = \frac{\langle \sigma v\rangle 
         m_2^3 x^2}{H} (f^2(x)-f_{eq}^2(x)) \,, 
\label{ecosmo1}
\end{equation}
where we have introduced the variables
\begin{equation}
  f(x) \equiv \frac{n(x)}{s_E}, \quad {\rm and} \quad     
  x \equiv \frac{s_E^{1/3}}{m_2},                                                          
\end{equation}
where $m_2$ is the WIMP mass and $s_E(T)$ is the thermal entropy density 
at the temperature $T$. $H(T) = (8\pi\rho(T)/3M_{\rm Pl}^2)^{1/2}$ is the Hubble parameter and $\langle \sigma v\rangle$ is the average WIMP annihilation rate which expression is defined below in Eq.~(\ref{ecosmo3}). We assume the standard adiabatic expansion law for the universe and use the standard thermal integral expressions for $s_E$ and for $H(T)$. Freeze-out temperature for our WIMPs is typically $T\sim {\cal O}(1-10)\rm \; GeV$ and thus the uncertainties in $s_E$ related to the QCD phase transition do not affect our analysis. The present ratio of $N_2$-number-density to the entropy density $f(0)$ is solved numerically from Eq.~(\ref{ecosmo1}) after which the fractional density parameter $\Omega_{N_2}$ of the Majorana WIMPs follows from
\begin{equation}
\Omega_{N_2}\simeq 5.5 \times 10^{11} \frac{m_2}{{\rm TeV}} f(0)\,.
\label{ecosmo8}
\end{equation}
From Eq.~(\ref{ecosmo1}) one sees that the relic density $f(0)$ essentially depends on the ratio $\langle \sigma v\rangle/H$; the smaller the ratio, the less time the WIMPs can remain in thermal equilibrium and thus  the larger is their relic density. It can be shown that the dependence is in fact almost linear: $\Omega_{N_2} \sim H/\langle \sigma v\rangle$ (see {\em e.g.}~\cite{Enqvist:1988dt}). 
As we assume the standard expansion history of the universe so that the $H$ is known, the solution $f(0)$ is determined by the annihilation cross section $\langle \sigma v\rangle$.  

For the thermally averaged annihilation cross section we use the expression~\cite{Gondolo:1990dk}:
\begin{equation}
   \langle  \sigma v \rangle = 
       \frac{1}{8m_2^{4}TK^{2}_2(\frac{m_2}{T})}
      \int_{4m_2^2}^{\infty} ds
                        \sqrt{s}(s-4m_2^2)K_1(\frac{\sqrt{s}}{T})
                        \sigma_{\rm tot}(s)
\label{ecosmo3}
\end{equation}
where $K_i(y)$s are modified Bessel functions of the second kind and $s$ is the Mandelstam invariant. For the total cross section $\sigma_{\rm tot}$ we considered the $N_2\bar{N_2}$ annihilation to the final states including all open fermion, gauge boson and scalar channels 
\begin{equation}
N_2 \bar N_2 \rightarrow f\bar{f}, \; W^{+}W^{-}, \; ZZ,\; ZH^0 
\; {\rm and} \; H^0H^0 \,,
\label{eq:channels}
\end{equation}
Here $H^0$ is the effective light ``SM-like'' Higgs state appearing in the mass operators (\ref{scalar_fermion}) and (\ref{scalar_fermion2}).
We did not take into account the WIMP annihilations to technifermions. This is because the technifermions would give only a small contribution to the fermionic annihilation channel in particularly in the case of heavier WIMPs, of which we are not interested here.
Also in the WIMP mass Scenario II, the annihilations to scalars $S$ were omitted assuming that these scalars are heavy. 
The cross sections for each channel shown in~(\ref{eq:channels}) were calculated without further approximations and all $s$-integrals were solved numerically. For these computations the needed
WIMP-Higgs and WIMP-gauge bosons couplings were given in section~\ref{sec:couplings}.
We also assumed that the unstable heavier neutrino state $N_1$ has decayed before the $N_2$ freeze-out. Thus, the particles present during the freeze out are just the Standard Model particles and the annihilating WIMP.

Let us mention that the WIMP annihilation cross section to fermionic final states used in this work differs slightly from the one used in \cite{Kainulainen:2009rb} (Eq. 3.6). This is because here we took into account the full expression of the $Z$-boson propagator (in unitary gauge) when calculating this cross section, as in \cite{Kainulainen:2009rb} only the part of the propagator proportional to $g_{\mu \nu}$ was used. 
However, the new terms, following from the use of the full $Z$ propagator, have only minor impact on the cross section and thus on the final WIMP density in the WIMP mass range of our interest.
For other applications the new terms can be relevant and thus we give the corrected cross section in the Appendix. More details about the computation of the cross sections can be found from \cite{Kainulainen:2009rb}.

Before going to the results let us summarize shortly what parameters affect on the relic density: the relic density is controlled by the annihilation cross section, which scale is set by the mixing angle $\sin \theta$ and the mass of the WIMP $m_2$. The Higgs-boson mass $m_H$ and the WIMP mass scenario affects considerably to the relic density for WIMP masses $m_2 \approx m_H/2$. Also the phase $\rho_{12}$ and the mass $m_1$ have an impact, mostly through the Higgs couplings, on our results. However, the mass of the charged lepton $m_E$ affects only very weakly on the relic density analysis. A more detailed characterization of the effects of different parameters on our results is given in the next section.

\section{Results}
\label{sec:results}

\subsection{Invisible Decay Width}

\begin{figure}
\includegraphics[width=0.49\textwidth]{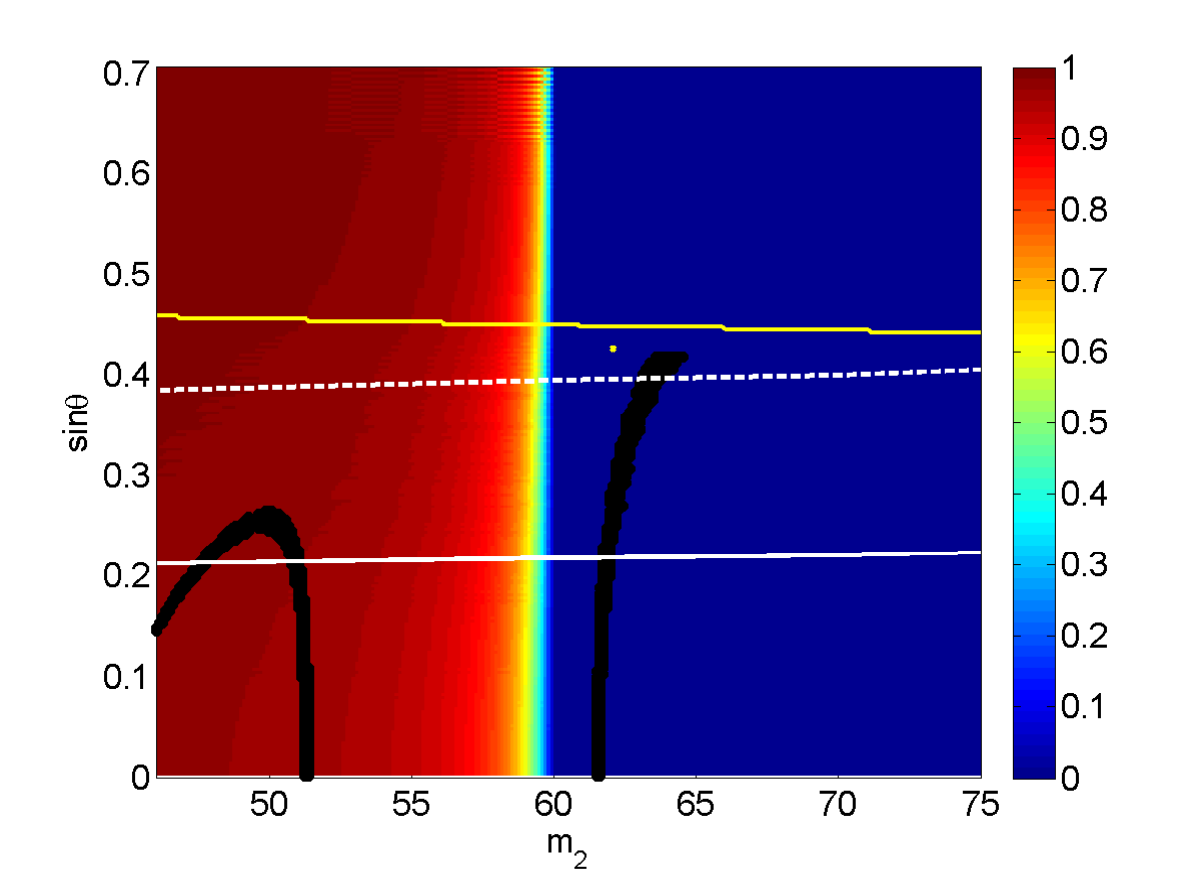}
\includegraphics[width=0.49\textwidth]{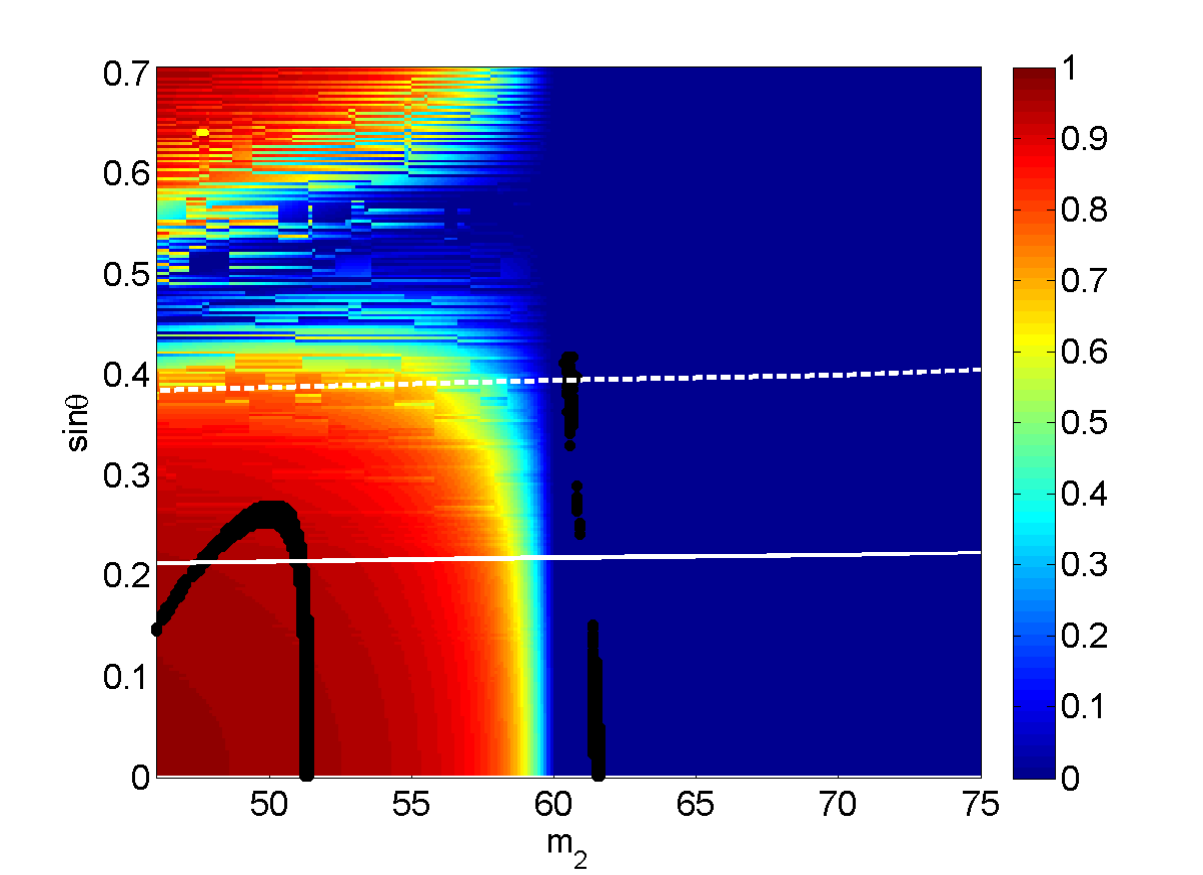}
\includegraphics[width=0.49\textwidth]{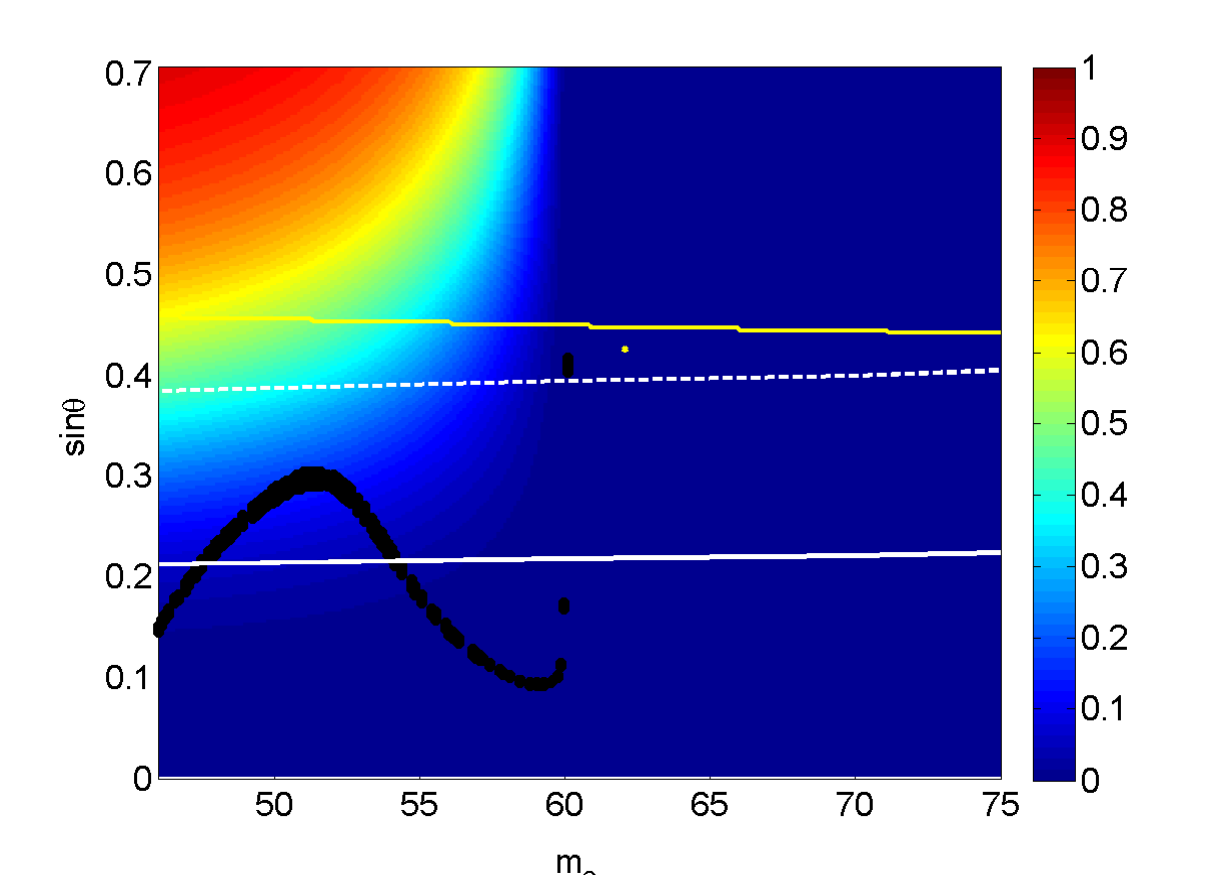}
\includegraphics[width=0.49\textwidth]{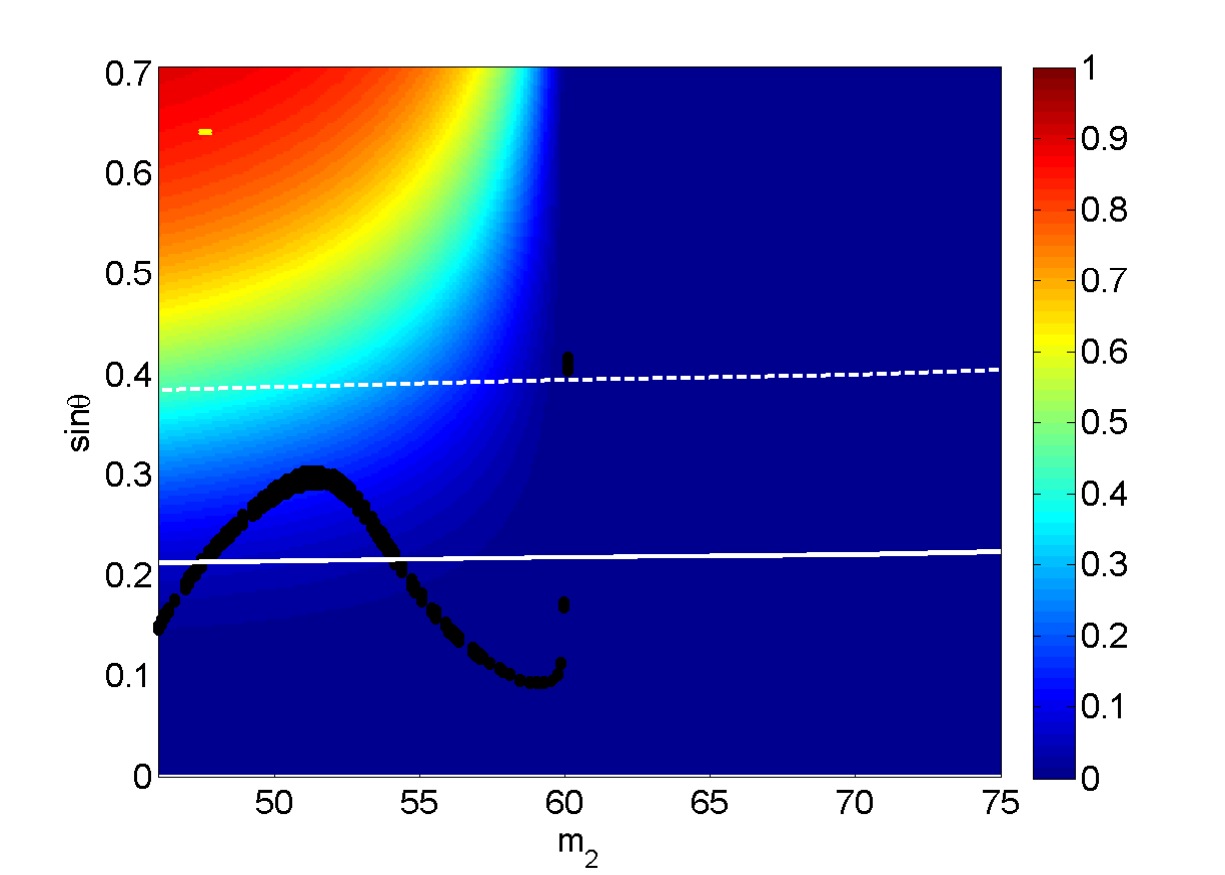}
\caption{The invisible branching ratio $R_\Gamma$ of the composite Higgs
 boson for $m_H=120$ GeV in scenario I (upper panel) and scenario II
 (lower panel). $\rho_{12}=-1(+1)$ in the left (right)
 panel. The black dots show points of the parameter space that produce
 acceptable relic density. The area above the white dashed line is ruled
 out by the XENON10-results, and the continuous white line shows the
 XENON100-limit, as discussed in section \ref{sec:cryolimits}. We do not show points of acceptable relic density above $\sin\theta > 0.42$, since these are in any case ruled out by the direct detection experiments. In the left panels the area
 above the topmost solid line (yellow), is ruled out by the oblique constraints.}
 \label{fig:mh120}
\end{figure}

\begin{figure}
\includegraphics[width=0.49\textwidth]{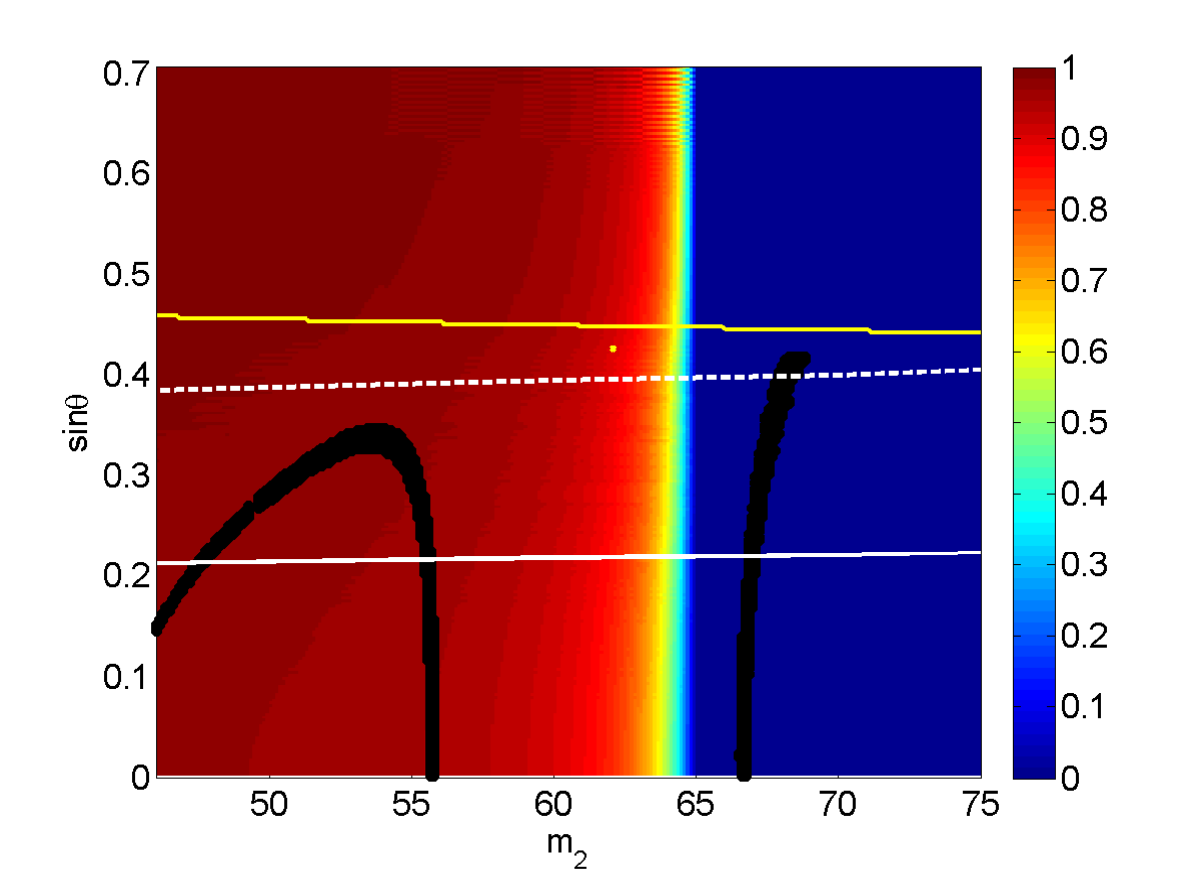}
\includegraphics[width=0.49\textwidth]{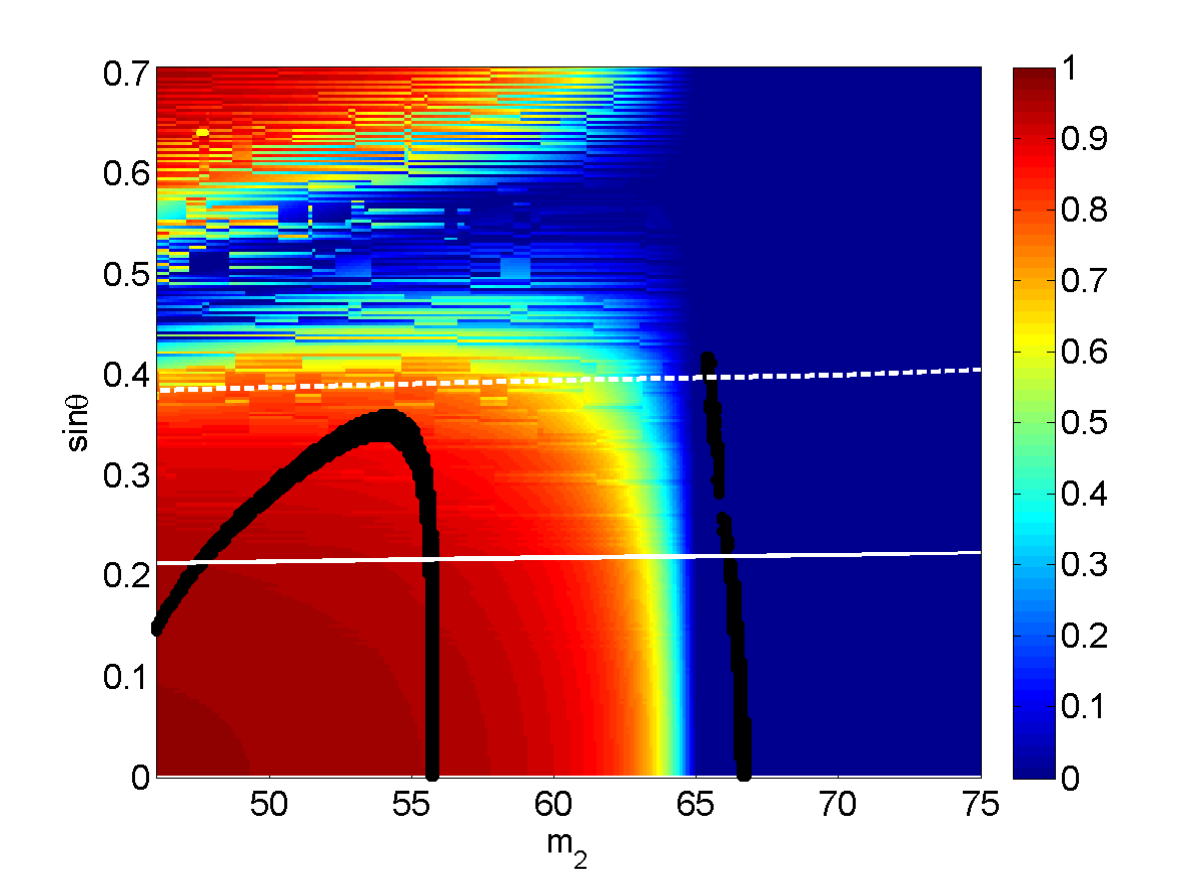}
\includegraphics[width=0.49\textwidth]{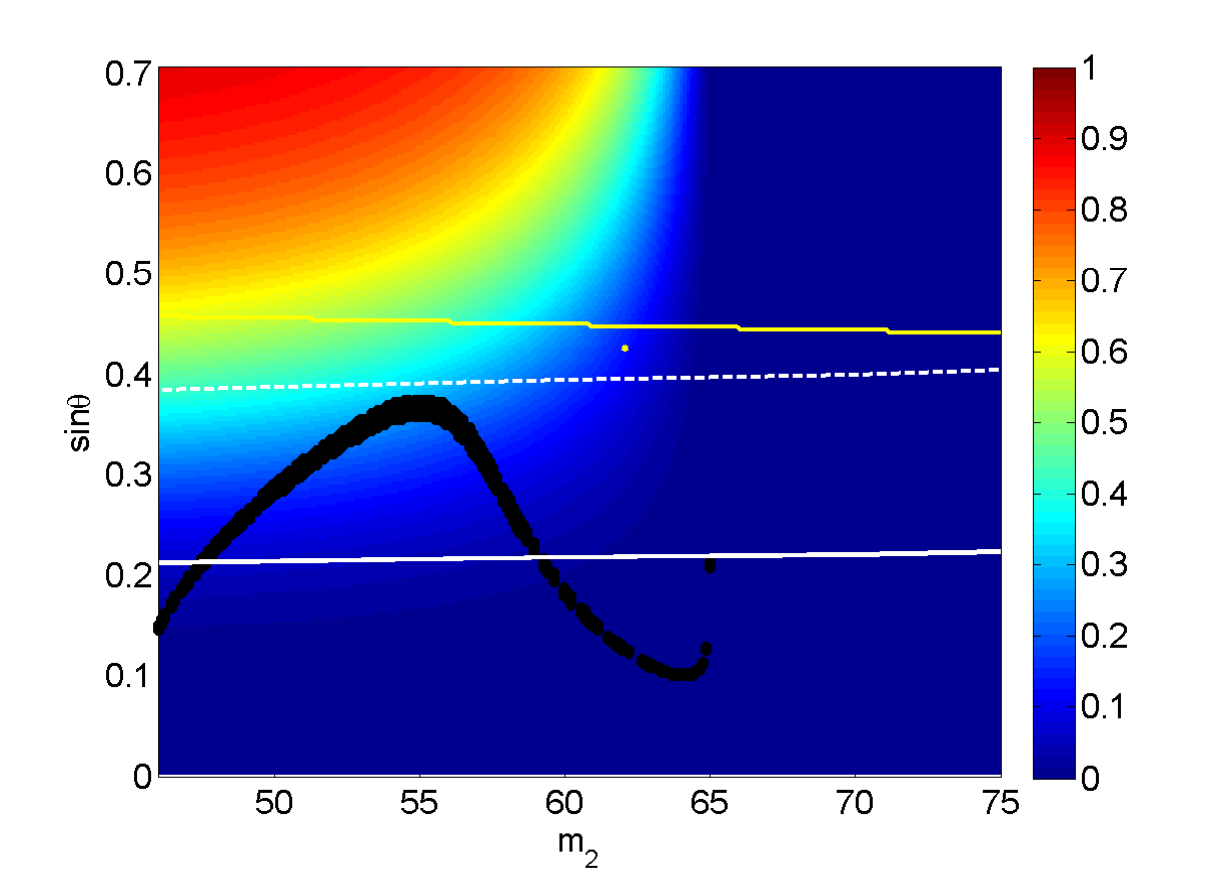}
\includegraphics[width=0.49\textwidth]{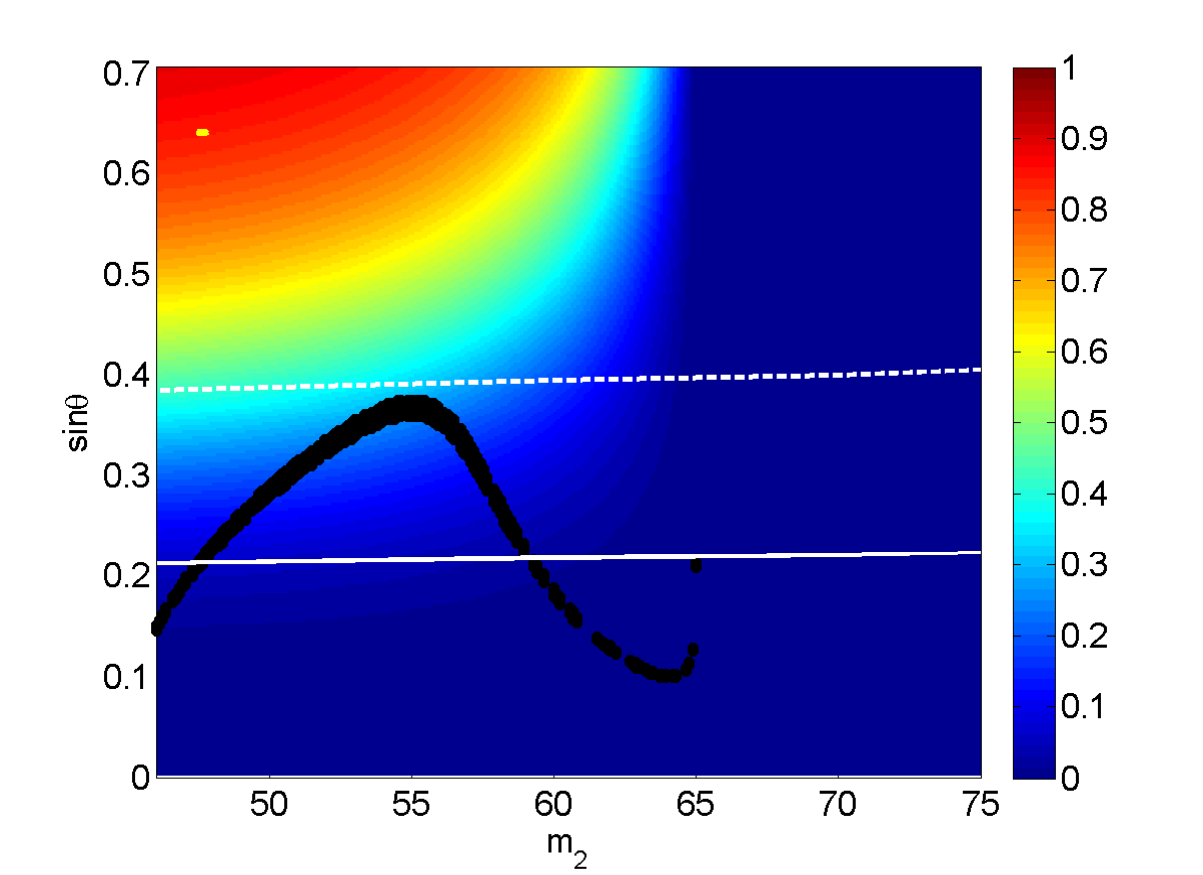}
\caption{Same as figure \ref{fig:mh120} but for $m_H=130$ GeV.}
\label{fig:mh130}
\end{figure}

\begin{figure}
\includegraphics[width=0.49\textwidth]{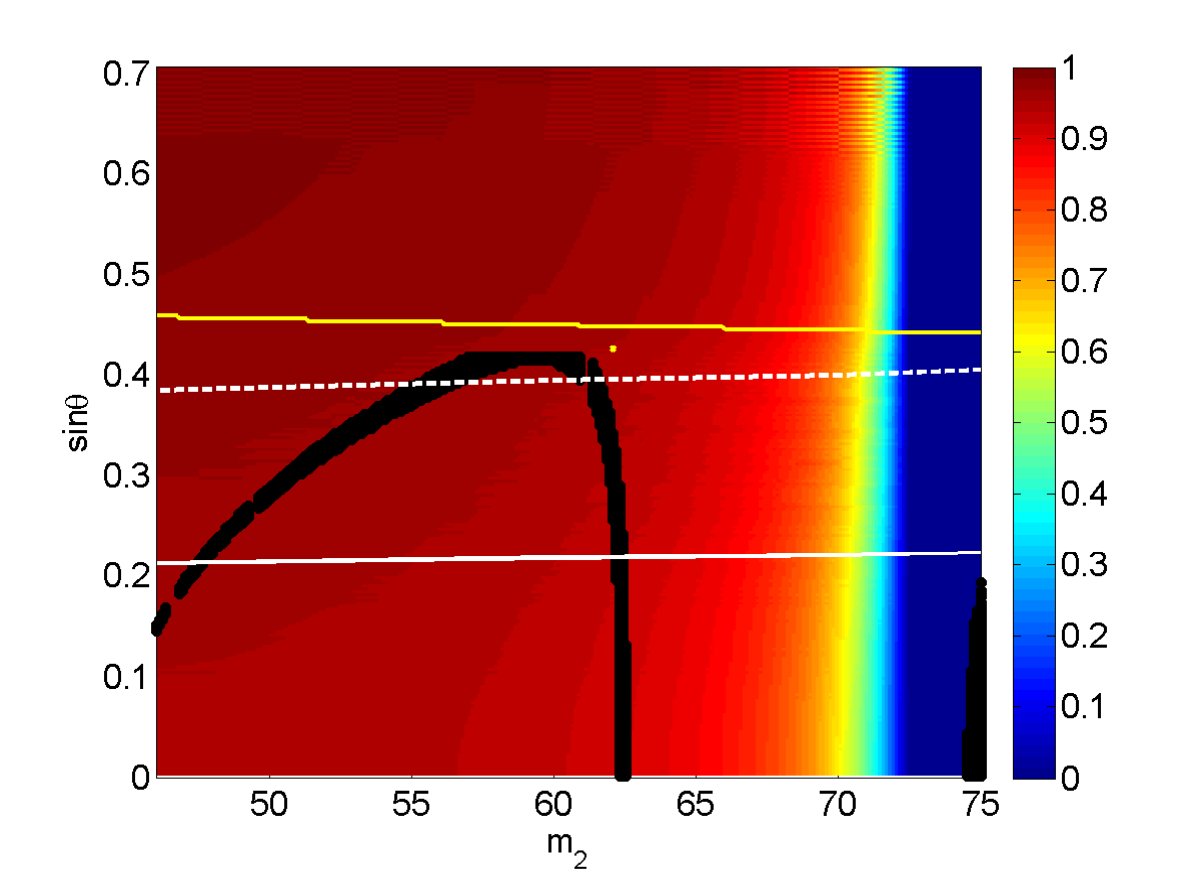}
\includegraphics[width=0.49\textwidth]{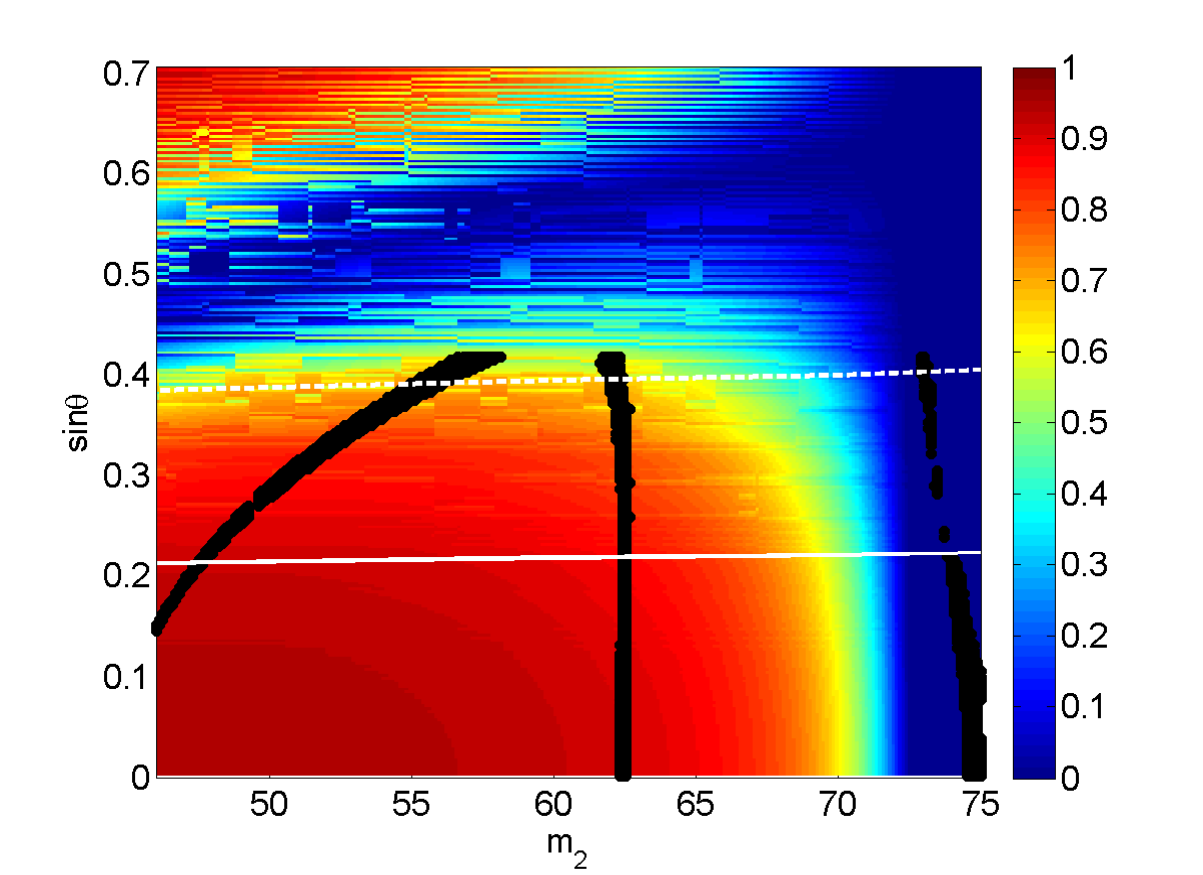}
\includegraphics[width=0.49\textwidth]{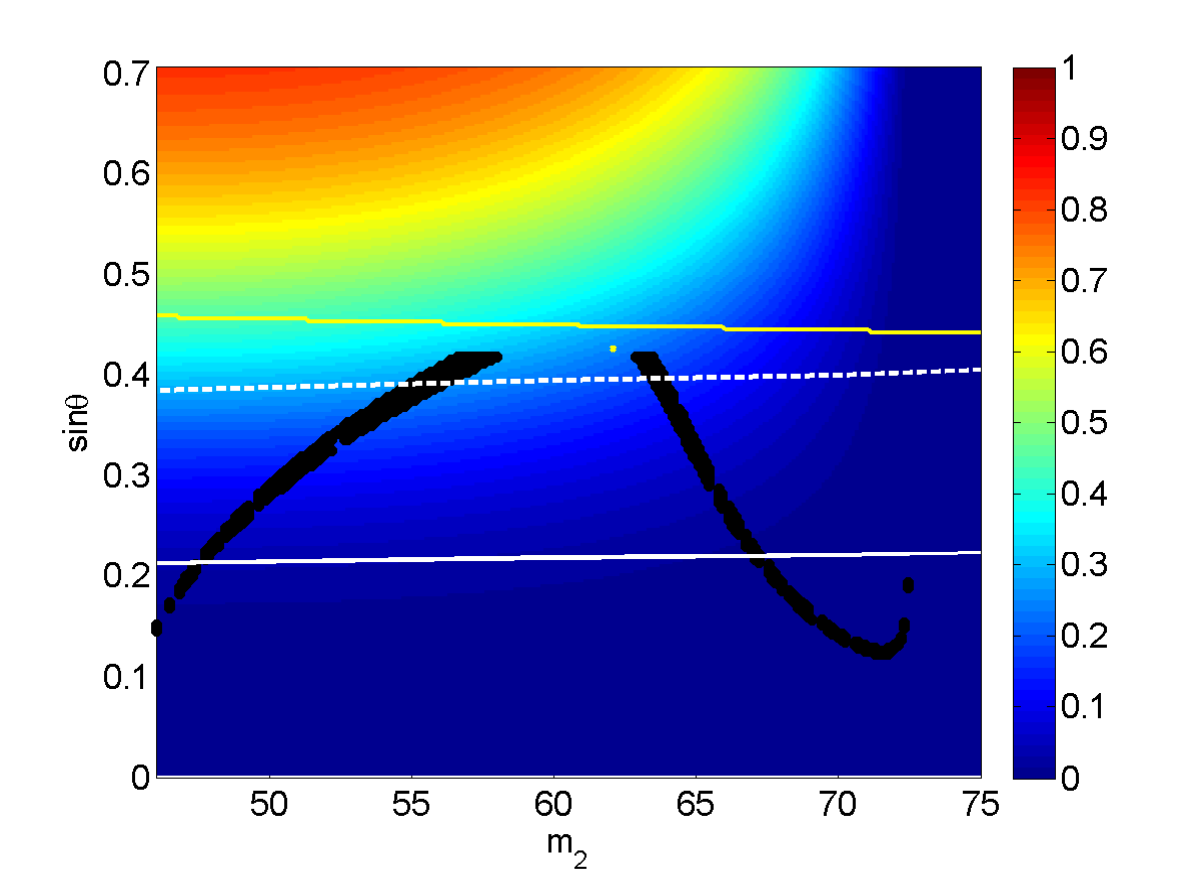}
\includegraphics[width=0.49\textwidth]{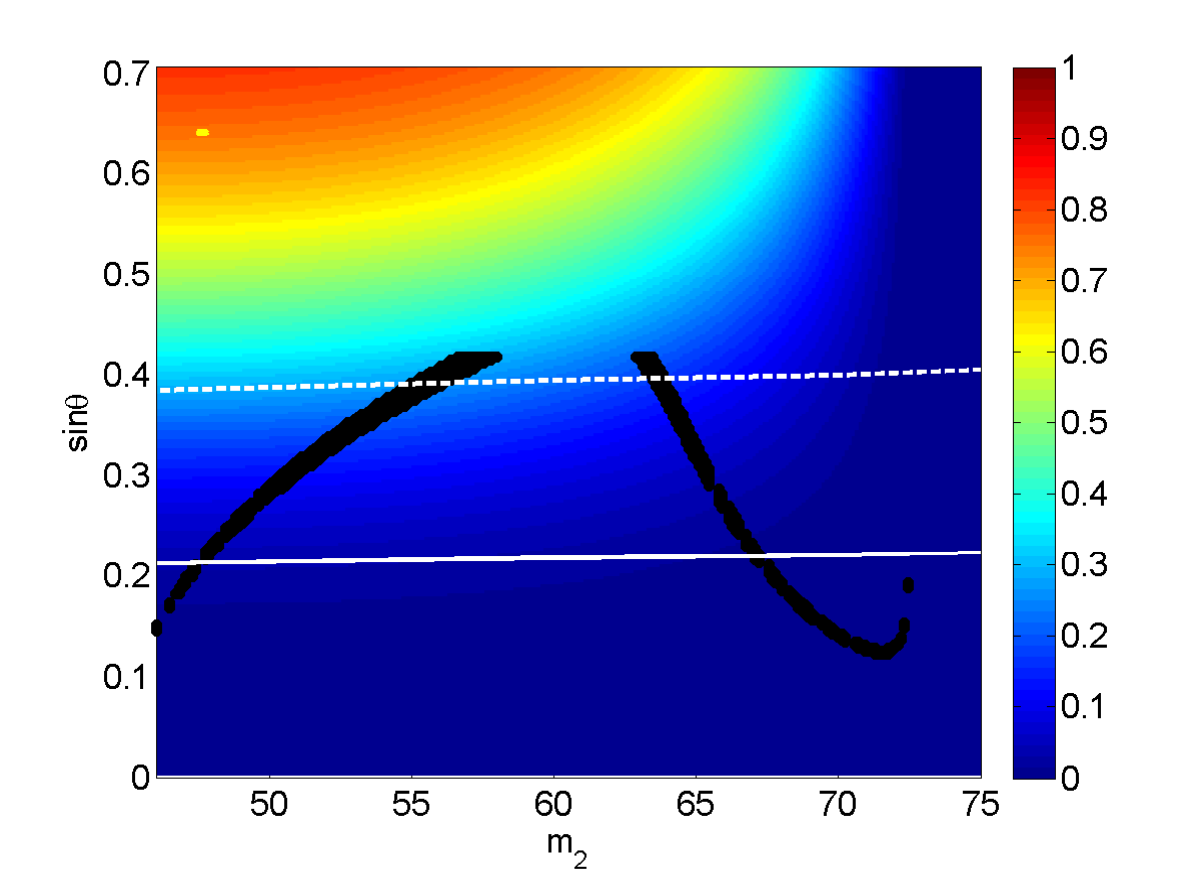}
\caption{Same as figure \ref{fig:mh120} but for $m_H=145$ GeV.}
\label{fig:mh145}
\end{figure}

As discussed in section \ref{sec:oblique} we do a scan over the
$(m_2,\sin\theta)$-plane and choose suitable values for $m_1$ and $m_E$
in each point of the plane, to meet the $(S,T)$-constraints. We then calculate $R_\Gamma$ and the relic density for each point. The decay width of the Higgs is computed at tree level, except for the $q\bar q$-channels where the leading logarithmic corrections are taken into account. The results are shown in figures
\ref{fig:mh120}, \ref{fig:mh130} and \ref{fig:mh145} for Higgs masses of
120 GeV, 130 GeV and 145 GeV, respectively. The colormap shows the value
of the invisible branching ratio $R_\Gamma$, and the black dots show the
points where the neutrino relic density has the correct value. The area
above the solid white line is ruled out by the earth based direct detection
dark matter searches.
\\

We shall now characterize the relic density results shown in Figs.~\ref{fig:mh120}-\ref{fig:mh145} 
more carefully.
As the WIMP mass $m_2$ and the mixing angle $\sin \theta$  set the scale for $\langle \sigma v\rangle$ these parameters also have the strongest impact on the final WIMP abundance $\Omega_{N_2} $.
For this reason we have projected all the suitable model parameter sets, which produce the measured DM 
density $\Omega_{DM}  \approx 0.19-0.23$~\cite{Nakamura:2010zzi} (consistent with combined 
WMAP7+H0 results \cite{WMAP7}), in the $(m_2, \sin \theta)$ plane.
Our results are also sensitive on the mass of the light composite Higgs particle $m_H$, especially in the 
WIMP mass region $m_2 \sim m_H/2$, where the WIMP annihilation cross section gets enhanced due to 
the increase of Higgs s-channel process at the Higgs pole. Our results also depend on the phase $
\rho_{12}$ which can be seen from the WIMP-Higgs interaction terms in Table~\ref{c_table}. 
Finally, the mass of the new charged lepton state, $m_E$, has only a subleading effect on the annihilation 
cross section and hence on the relic density.  Indeed, the charged state contributes to the cross section 
only as a virtual state in the $t$-channel process in WIMP annihilations into $W^+W^-$ final states.
In particular, for WIMP masses $m_2< M_W$ of which we are interested here, this annihilation channel 
has only very small effect in the cross section integral $\langle \sigma v \rangle$ given in Eq. 
(\ref{ecosmo3}). In principle similar arguments, for the subleading effects in annihilation cross section, 
holds also for the heavier neutral state $N_1$ with mass $m_1$, as also $N_1$ contributes only as virtual 
state in the $t$-channel annihilations into $Z H^0$ and $H^0 H^0$ final states. 
However, for WIMP mass Scenario I, as the Higgs couplings given in Table~\ref{c_table}  include the 
mass ratio of the neutral states $m_1/m_2$, also $m_1$ will affect the annihilation cross section and thus 
the final WIMP abundance for the WIMP masses $m_2 \sim m_H/2$.

For WIMP masses $m_2< M_W$ the WIMP annihilation cross section is determined by the WIMP 
annihilations to SM fermions via $Z$-boson in $s$-channel process. 
The annihilation cross section in this case is directly proportional to mixing angle: $\langle  \sigma v 
\rangle \propto \sin^4 \theta $ (this can be immediately realized from the $Z$-WIMP interaction term given 
in Eq.~(\ref{eq:intlagrange})).
Now for a fixed mixing angle the cross sections gets enhanced at the $Z$-pole. As the DM density is 
inversely proportional to the annihilation cross section, the DM density will in principle decreases in this 
case. Thus, to keep the $\Omega_{N_2} $ constant, indicating that the cross section is constant, we need 
to decrease the value of the mixing angle to compensate for the effect of the $Z$-pole. 
This produces a dip on the suitable mixing angle values for WIMP masses $m_2 \sim M_Z / 2$.

Similarly the effect of the Higgs resonance can be seen in Figs. \ref{fig:mh120} -\ref{fig:mh145} as a dip in 
the mixing angle values for the WIMP masses $m_2 \rightarrow m_H/2$. 
For the WIMP mass Scenario I the effect of the Higgs pole is more dramatic than the effect of the $Z$-pole. 
This is so because in the Higgs case the cross section depends only weakly on 
the mixing angle and thus the mixing angle is not able to sufficiently compensate for the pole. Indeed, for 
WIMP mass 
values $m_2 \approx m_H/2$ the mixing angle suppression is insufficient to keep the cross section small 
enough for production of the correct relic density. Thus the dip in the suitable mixing angle values for the 
WIMP masses $m_2 \rightarrow m_H/2$ becomes very deep and eventually it cuts the suitable parameter 
space well before $m_2$ reaches the value $m_H/2$. However, for Scenario II the cross section is again 
proportional to the mixing angle, as can be realized from the WIMP-Higgs coupling factor $C_{22}^h$ 
given in Table~\ref{c_table}, and thus the Higgs pole can be compensated for by decreasing the mixing 
angle and the correct relic density can be obtained for larger range of $m_2, \sin \theta$ parameter points. Let us also mention that in Scenario II the similarity of the DM density curves in different $\rho_{12}$ cases follows from the fact that, as the $C^h_{22}$ factor is independent of the $\rho_{12}$ parameter in this case, the $\langle \sigma v \rangle$ for light WIMP masses $(m_2 < m_Z)$ is effectively independent of the $\rho_{12}$ factor.

In all our plots we have set an upper limit for the mixing angle $\sin \theta <$ 0.42. This cuts the suitable DM density regions/curves
for heavy WIMP masses ($m_2 \gtrsim m_H/2$) e.g. in lower panels of Fig. \ref{fig:mh120} DM density curve is cut for WIMP masses larger than $\sim 60$ GeV. Similar cuts can be realized in other figures. There are two reasons for setting this cut/upper bound. 
Firstly, we are interested on relatively light WIMPs i.e. 45 GeV $< m_2 < $ 75 GeV, and in this mass region $\sin \theta$ values larger than $\approx 0.4$ are excluded already by the null results of direct DM search XENON10 experiment, as is explained in Sec.~\ref{sec:cryolimits} and shown in the Figs. \ref{fig:mh120} -\ref{fig:mh145} as the white dashed line. Thus, we do not need to plot the curves for heavier DM masses. Moreover, the model results for heavier WIMP masses were already shown in \cite{Kainulainen:2009rb}.
Secondly, in this work we are particularly interested in the Higgs decays into invisible DM channel, and since this decay is kinematically allowed only for $m_2 \leq m_H/2$, in this case plotting the DM density curve for much larger DM mass values than $\approx m_H/2$ is not interesting. 
Saying all this, one should still keep in mind that in principle the DM density curve continues in all figures to the right (as shown in \cite{Kainulainen:2009rb}), and that for large DM masses there still might be suitable parameters space left i.e. $(m_2, \sin \theta)$ points which produce correct DM density and are consistent with the EW precision measurements and are not excluded by the DM search experiments.

\subsection{Cryogenic limits}
\label{sec:cryolimits}

Here we summarize and update the strongest observational constraints for our WIMP candidate following 
from the direct DM detection experiments XENON10-100
~\cite{Angle:2007, Angle:2008we, Aprile:2011hi}. Part of these constraints were already shown in the previous result section, and here we 
demonstrate how these constraints were derived and also analyze the constraints set by these 
experiments in more detail. Other weaker constraints arising from LEP measurement of Z-decay width and from CDMS
~\cite{Ahmed:2008eu} were studied in \cite{Kainulainen:2009rb}, and constraint coming from indirect DM detection in particularly from neutrino detectors like Super-Kamiokande were studied in detail~\cite{Belotsky:2008vh}; see also indicative limits given in \cite{Kainulainen:2009rb}
 \footnote{ IceCube and SuperKamiokande collaborations have updated their constraints for spin-dependent WIMP-nucleon cross sections in \cite{IceCube:2011aj}
and \cite{Tanaka:2011uf} respectively. When comparing these new constraints with the older SUSY WIMP constraints given in~\cite{Desai:2004pq} and~\cite{Abbasi:2009uz} by Super-Kamionkande and IceCube collaborations, respectively, one notices that the WIMP-nucleon cross section especially for heavier WIMP masses ($m \gsim 100$ GeV) is now more constrained. These constraints are model dependent and are given for SUSY models. Thus they can be considered only as indicative limits for our model. In the analysis done for our model in~\cite{Belotsky:2008vh} and \cite{Kainulainen:2009rb}, the same Super-Kamiokande data as in the older SUSY analysis in~\cite{Desai:2004pq} has been used. As an outcome, only WIMP masses between $\sim 100 - 200$ GeV in our case were excluded, and one can conclude that in the WIMP mass range ($\sim 45 - 75$) GeV, which we consider here, these new limits are less severe for our model when compared to the XENON100 limits given below.}.
Further, the limits following from FERMI-LAT gamma ray data for this model will be studied elsewhere \cite{cp3col}.

We can set constraints for our model using both {\em spin-dependent} (SD) and {\em spin-independent} 
(SI) WIMP-nucleon interactions limits given by XENON10-0 collaborations.
For our model the  spin-dependent WIMP-nucleon interactions proceed via $Z$-boson exchange, while 
the spin-independent interaction is Higgs mediated.
As demonstrated in \cite{Kainulainen:2009rb},  for Higgs masses $m_H > 200$ GeV, constraints arising 
from spin-dependent  WIMP-nucleon cross section limits were more stringent than those following from 
the spin-independent  cross section limits. However, here we consider lighter Higgs masses and as we 
are at the same time interested on relatively light WIMPs, the constraints arising from the spin-
independent 
WIMP-nucleon interactions become stronger than the ones following from spin-dependent WIMP-nucleon 
interactions. It actually turns out that the WIMP mass Scenario I is basically excluded by the spin-
independent limits (in the case of light Higgs masses). This is because the WIMP-Higgs coupling factor 
$C_{22}^h$ is not suppressed by the mixing angle, but is of the order of the standard $4^{\rm th}$ family 
Majorana neutrino-Higgs couplings, which are already excluded by the DM direct detection experiments~\cite{Angle:2008we}. 
However, for Scenario II the WIMP-Higgs interaction is suppressed with the mixing angle as $C_{22}^h = 
\sin^2 \theta$. Thus in Scenario II the WIMPs can avoid being ruled out by the spin-independent limits, 
and actually for this scenario the spin-dependent limits give more stringent constraints for the model.

We start by deriving the constraint following from the SD cross section limits, as these were already shown 
in the previous result section. The best current spin-dependent cross section limit comes from the 
cryogenic dark matter search XENON10 experiment~\cite{Angle:2008we}. This experiment has given their 
(spin-dependent) constraints  for a standard model 4$^{\rm{th}}$ family Majorana neutrino in reference~
\cite{Angle:2008we}  and explicitly plots the expected count rate in their detector for this case. 
Now, the (SD) count rate $N \propto \sigma_0$ for our WIMP,  differs from the standard model case only 
by a simple scaling of the cross section factor $\sigma_0$:
$ \sigma_{0,\rm SM} \rightarrow \sigma_{0,\rm Mix} = \sin^4\theta \sigma_{0,\rm SM}$, where $\sigma_0 $ 
accounts for the spin-dependent WIMP-nucleus cross section at the zero momentum transfer limit. Using 
this information we can convert the XENON10 results for a 4$^{\rm{th}}$ family SM-neutrino to an upper 
limit on the mixing angle as a function of mass:
\begin{equation}
\sin \theta (m_2)< \left(\frac{N_{\rm{limit}}(m_2)}{N_{\rm{SM}}(m_2)}\right)^{1/4}.
\label{eq:uppersin}
\end{equation}
The function $N_{\rm{SM}}(m_2)$ was read from the left panel in Fig.2 of ref.~\cite{Angle:2008we} and the 
function $N_{\rm{limit}}(m_2)$ was approximated by a linear interpolation between the values of 
$N_{\rm{SM}}(m_2)$ at the high and low mass ends of the SM-exclusion region in the same figure. The 
white dashed line in Figs.~\ref{fig:mh120}-\ref{fig:mh145}  show the excluded region corresponding to the 
upper limit (\ref{eq:uppersin}). 
To make this constraint even tighter we project this limit to correspond to the limit following from the null 
results of XENON100 experiment. This is done by simply scaling the above limit for the mixing angle  
(\ref{eq:uppersin}) with the fourth root of the ratios of the mass-day exposures of the two experiments: $\sin
\theta_{\rm lim} \rightarrow \sin\theta_{\rm lim} (E_{Xe10}/E_{Xe100})^{1/4}$. We use the full exposure 
$E_{Xe10}=136$kg-days for the XENON10 experiment and exposure $E_{Xe100}=1471$kg-days for the 
XENON100, which was reported by the XENON100 collaboration in connection with their April 2011 
results in \cite{Aprile:2011hi}. This tighter limit corresponds to the white solid line in Figs.~\ref{fig:mh120}-
\ref{fig:mh145}.

As already mentioned above, our WIMP has also spin-independent interactions mediated by the 
Higgs field, which were not accounted for in the treatment leading to the constraint (\ref{eq:uppersin}).
The spin-independent limits from XENON10-100 are given in~\cite{Angle:2007} and \cite{Aprile:2011hi} 
respectively, in terms of the WIMP-nucleon cross section. In our model this cross section (in the zero 
momentum transfer limit) is given by
\begin{equation}
 \sigma_{0}^{\rm SI, n} =  (C_{22}^h)^2 \,\frac{8 G_F^2\mu_{\rm n}^2}{\pi} \frac{m_2^2 m_{\rm n}^2}{m_H^4} f^2,
\label{cosmo14}
\end{equation}
where $n$ refers to a nucleon, $\mu_{\rm n}$ is the WIMP-nucleon reduced mass and $f$ is the Higgs 
nucleon coupling factor accounting for the quark scalar currents in the nucleons
\footnote{We use rather conservative value $f=0.5$ in our computations. However, the uncertainties in $f$ 
are pretty large, following mostly from the uncertainties in the value of the pion nucleon sigma term (See 
e.g. \cite{Ellis:2008hf,Alarcon:2011zs,Bali:2012rs,Cheng:2012qr}
). The values of $f$ may vary from $\approx 
0.3$ up to $\approx0.6$.}.

In the figures we have plotted $\sigma_{0}^{\rm SI, n}$  from (\ref{cosmo14}) for our model using those 
$m_2, \sin \theta$ and $ m_1$ values in $C_{22}^h$ which produce the correct DM density 
$\Omega_{N_2}$. Thus each point in ($ \sigma_{0}^{\rm SI, n}, m_2$) plane can be mapped to a point in 
one of the Figs.~\ref{fig:mh120}-\ref{fig:mh145}  naturally respecting the $m_H$ and $\rho_{12}$ cases.  
Also the exclusion curves following from XENON10~\cite{Angle:2007} (black dashed), CDMS~
\cite{Ahmed:2008eu} (black dash-dotted), XENON100~\cite{Aprile:2011hi} (black solid) results and 
XENON100 projected (black dotted) are plotted in the figures. The colormap shows the value of the Higgs 
branching ratio $R_\Gamma$ to the invisible DM sector.
\begin{figure}
\includegraphics[width=0.49\textwidth]{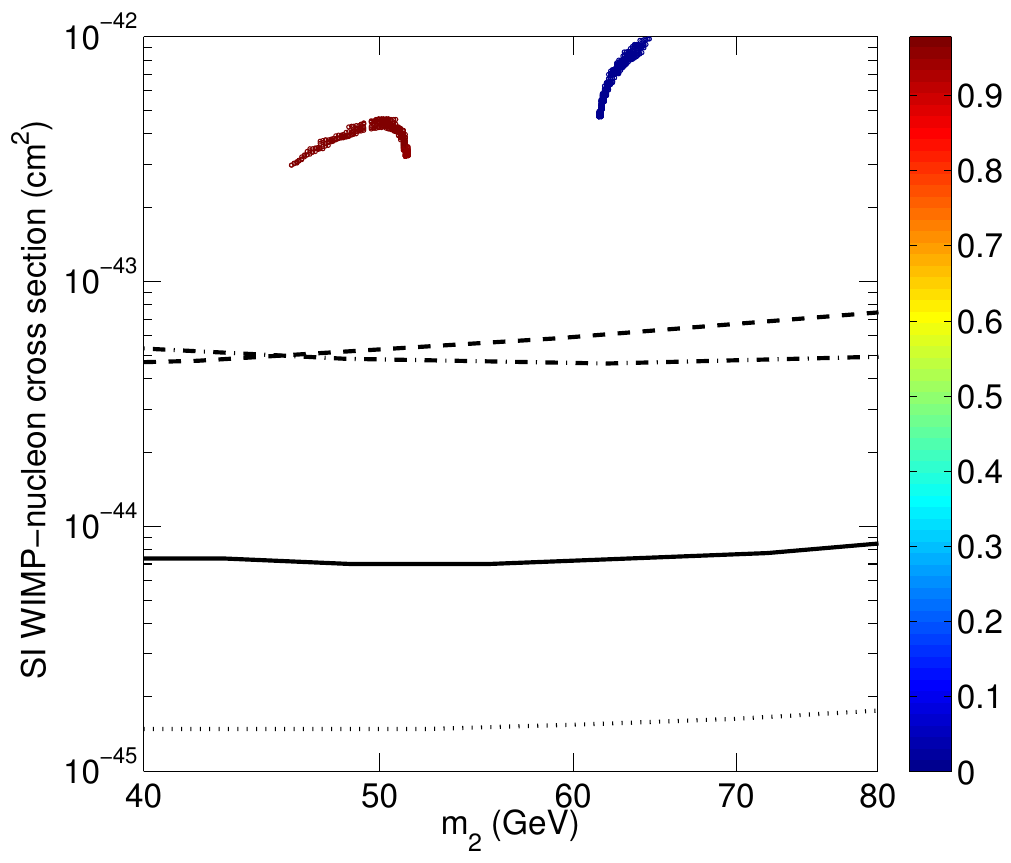}
\includegraphics[width=0.49\textwidth]{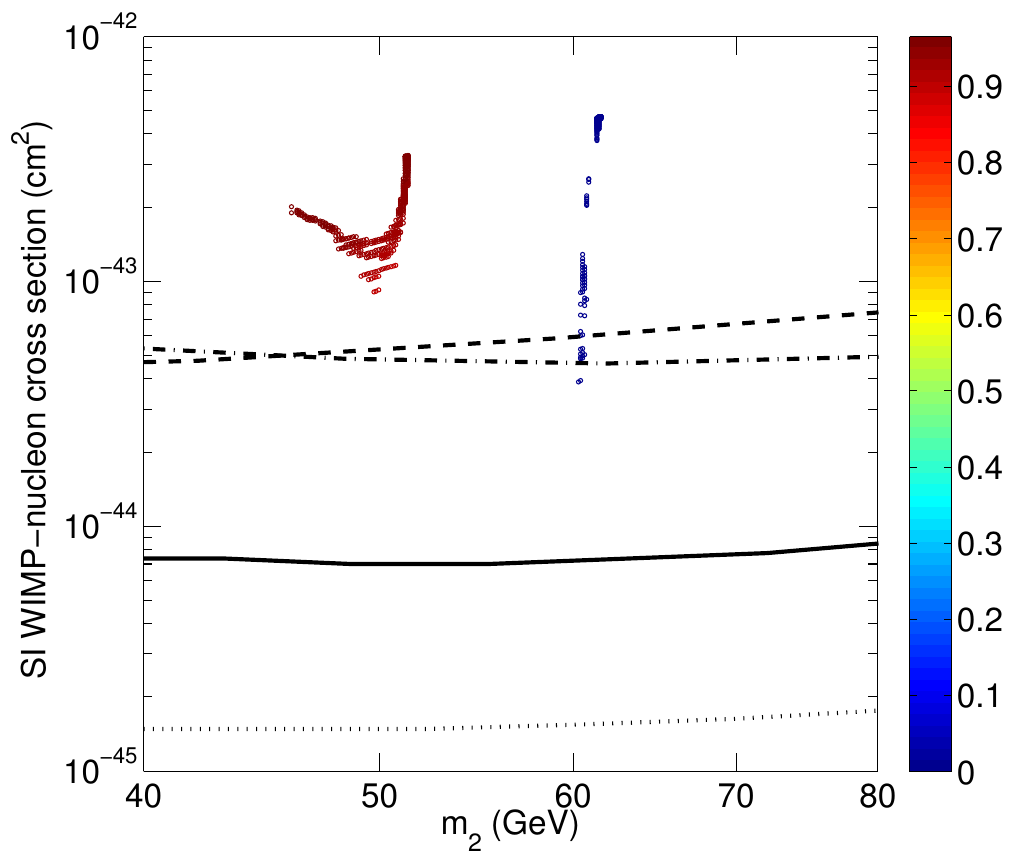}
\caption{Shown are the spin independent WIMP-nucleon cross sections in the case of $m_H=120$ GeV, in Scenario I. $\rho_{12}=+1(-1)$ in the left (right) plot. Colored patch shows the model predictions, yielding $\Omega_{N_2}$ between values 0.19-0.23, and the colormap indicates the value of the invisible branching ratio $R_\Gamma$. The black curves line the exclusion regions  
following from XENON10~\cite{Angle:2007} (dashed), CDMS~\cite{Ahmed:2008eu} (dash-dotted), and XENON100~\cite{Aprile:2011hi} (solid) experiments. The black dotted curve is the full expected XENON100 sensitivity. For production of the part of the experimental curves (XENON10, CDMS, XENON100 expected) shown here and in Figs.~\ref{fig:SImh130}-\ref{fig:SImh145} we used the tools of ref. \cite{dmtools}.}
\label{fig:SImh120}
\end{figure}
\begin{figure}
\includegraphics[width=0.49\textwidth]{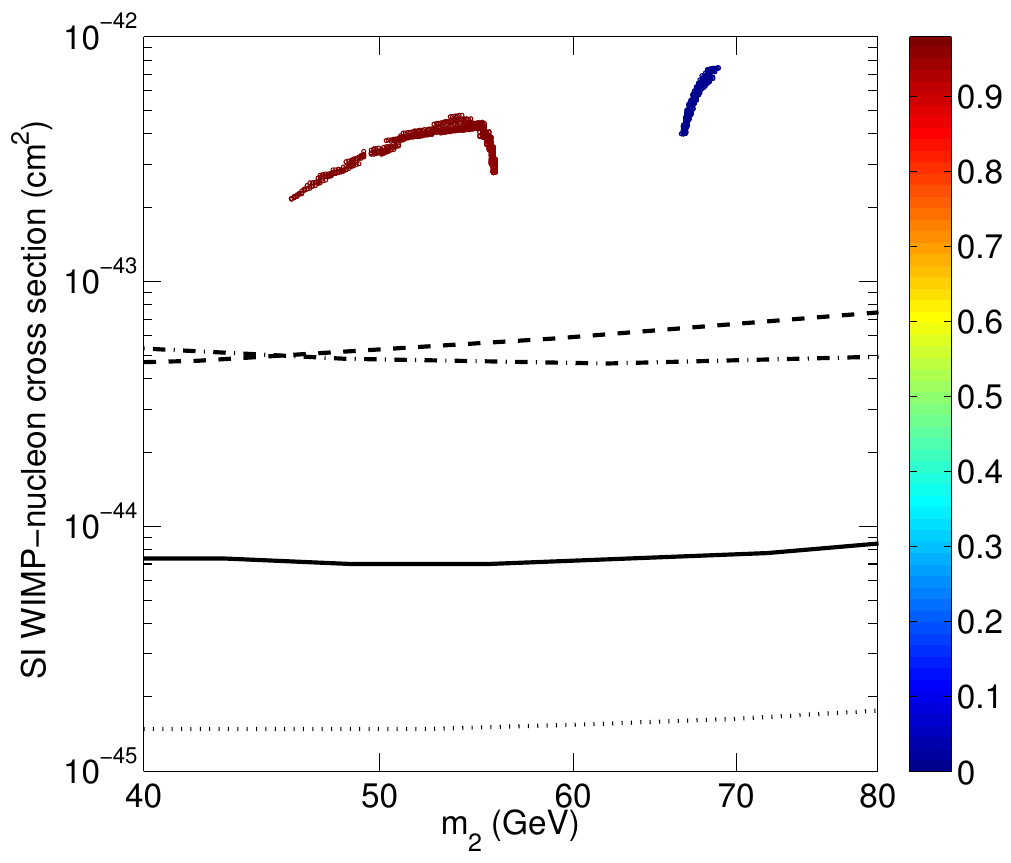}
\includegraphics[width=0.49\textwidth]{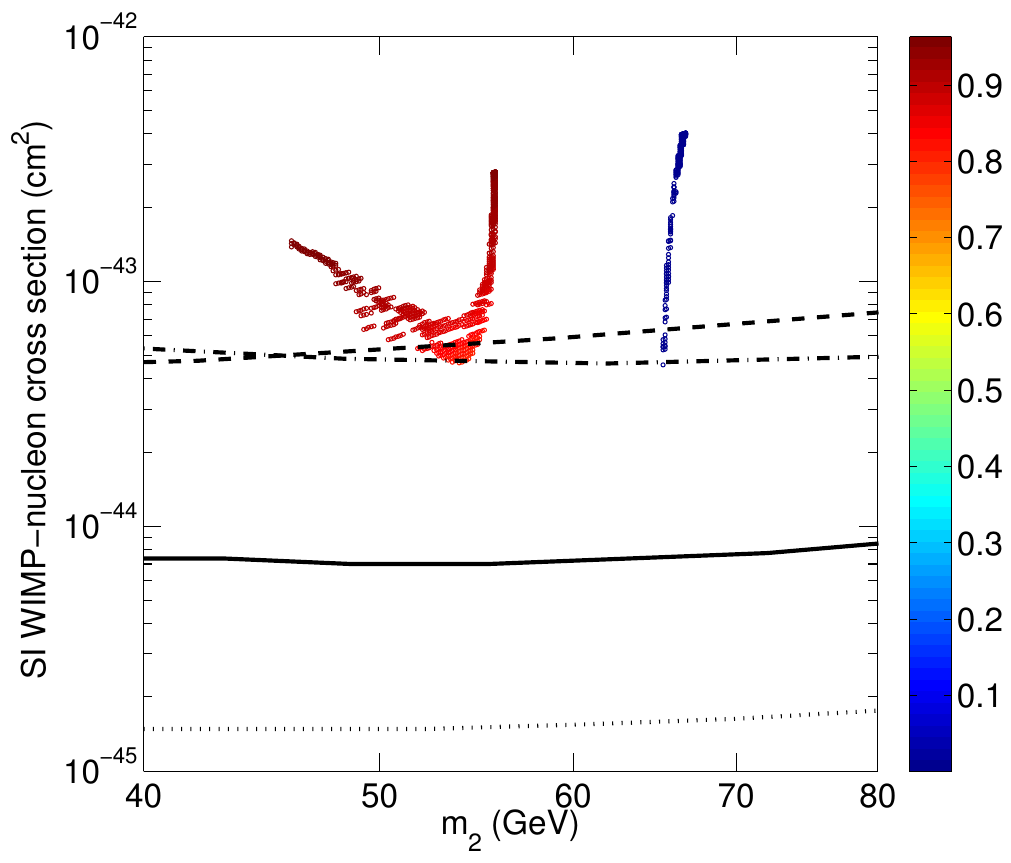}
\caption{Same as figure \ref{fig:SImh120} but for $m_H=130$ GeV.}
\label{fig:SImh130}
\end{figure}
\begin{figure}
\includegraphics[width=0.49\textwidth]{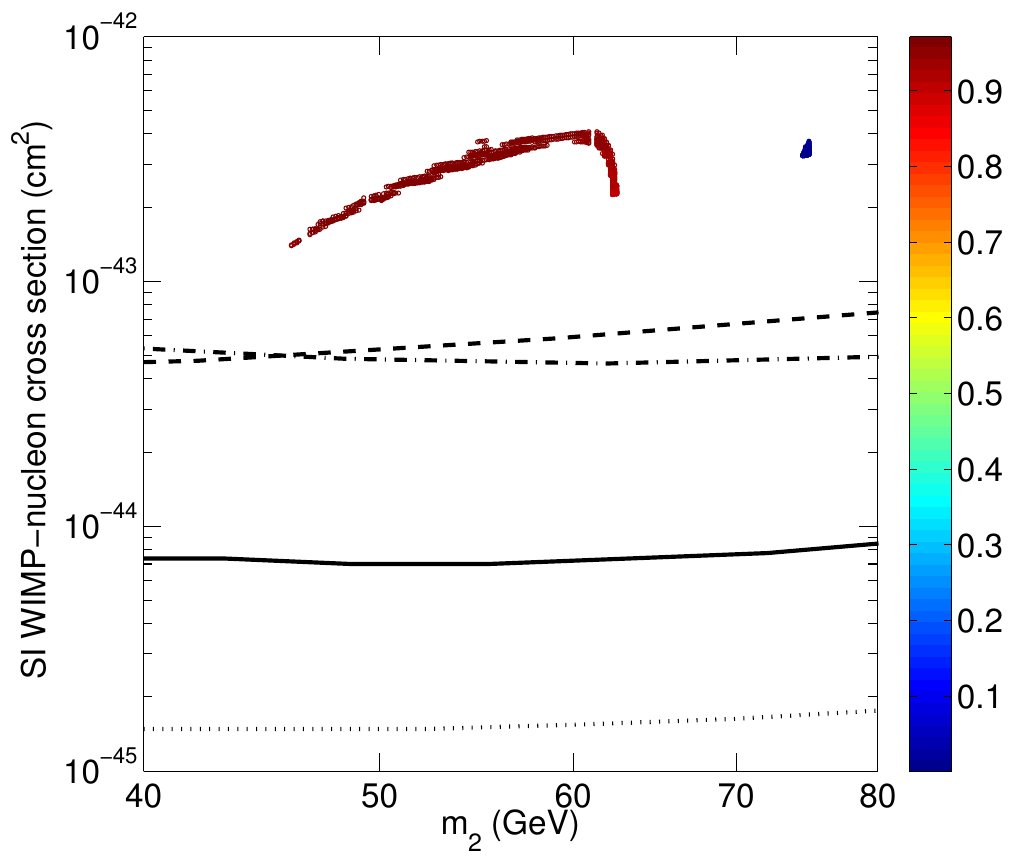}
\includegraphics[width=0.49\textwidth]{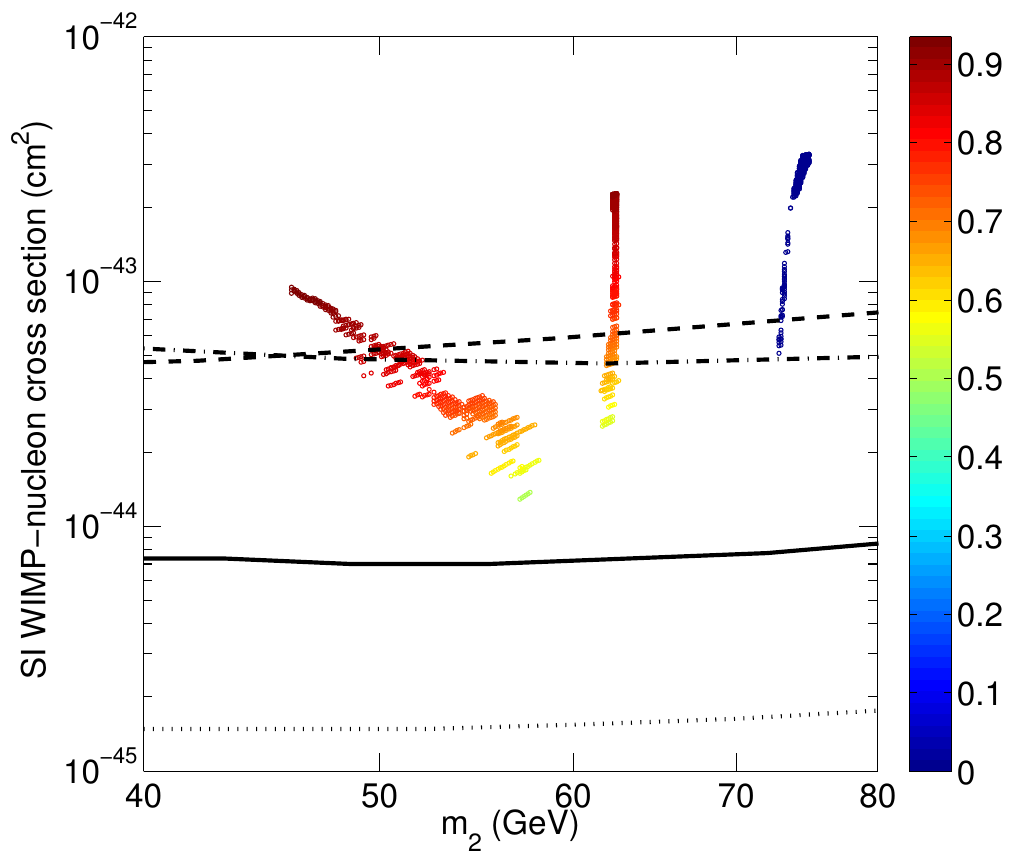}
\caption{Same as figure \ref{fig:SImh120} but for $m_H=145$ GeV.}
\label{fig:SImh145}
\end{figure}

\begin{figure}
\includegraphics[width=0.49\textwidth]{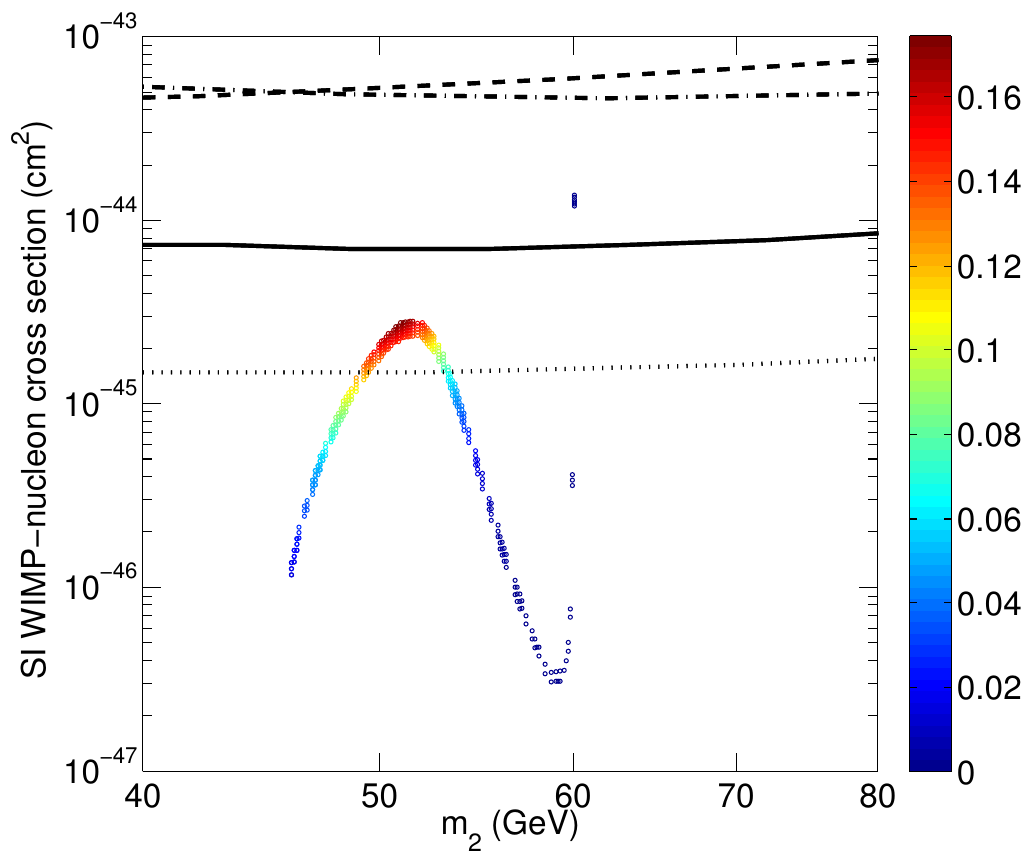}
\includegraphics[width=0.49\textwidth]{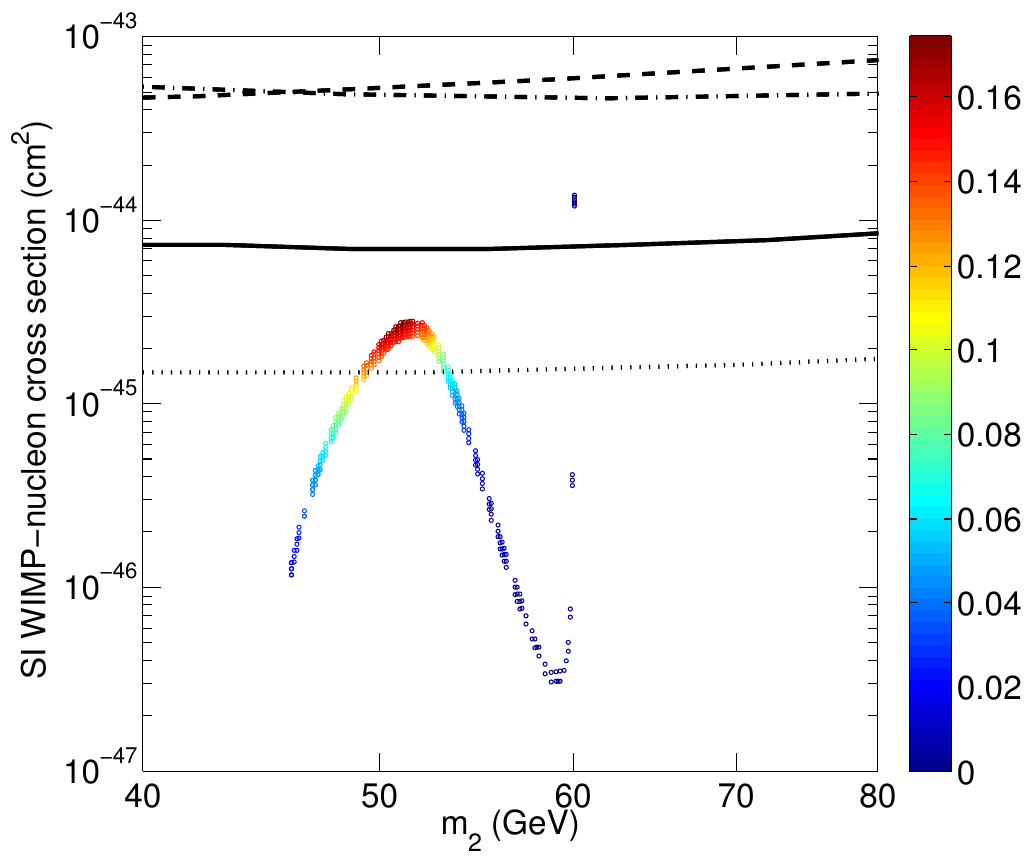}
\caption{Same as figure \ref{fig:SImh120} but for Scenario II, with $m_H=120$ GeV.}
\label{fig:SImh120S2}
\end{figure}
\begin{figure}
\includegraphics[width=0.49\textwidth]{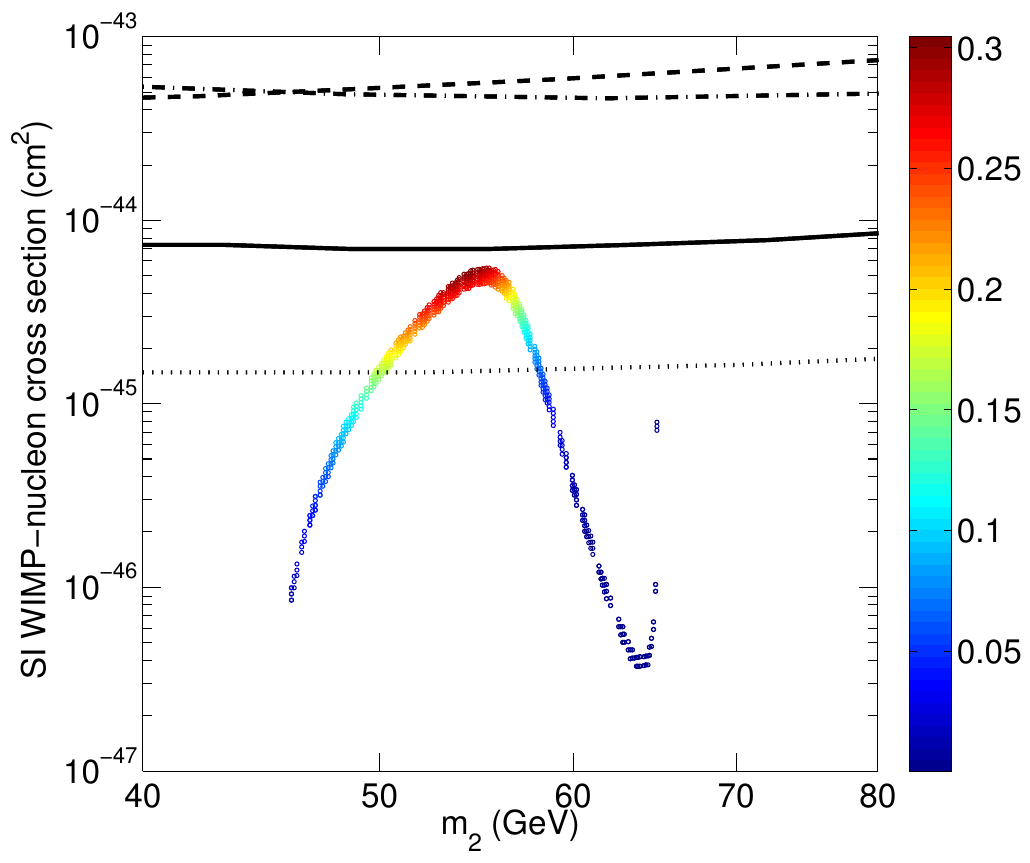}
\includegraphics[width=0.49\textwidth]{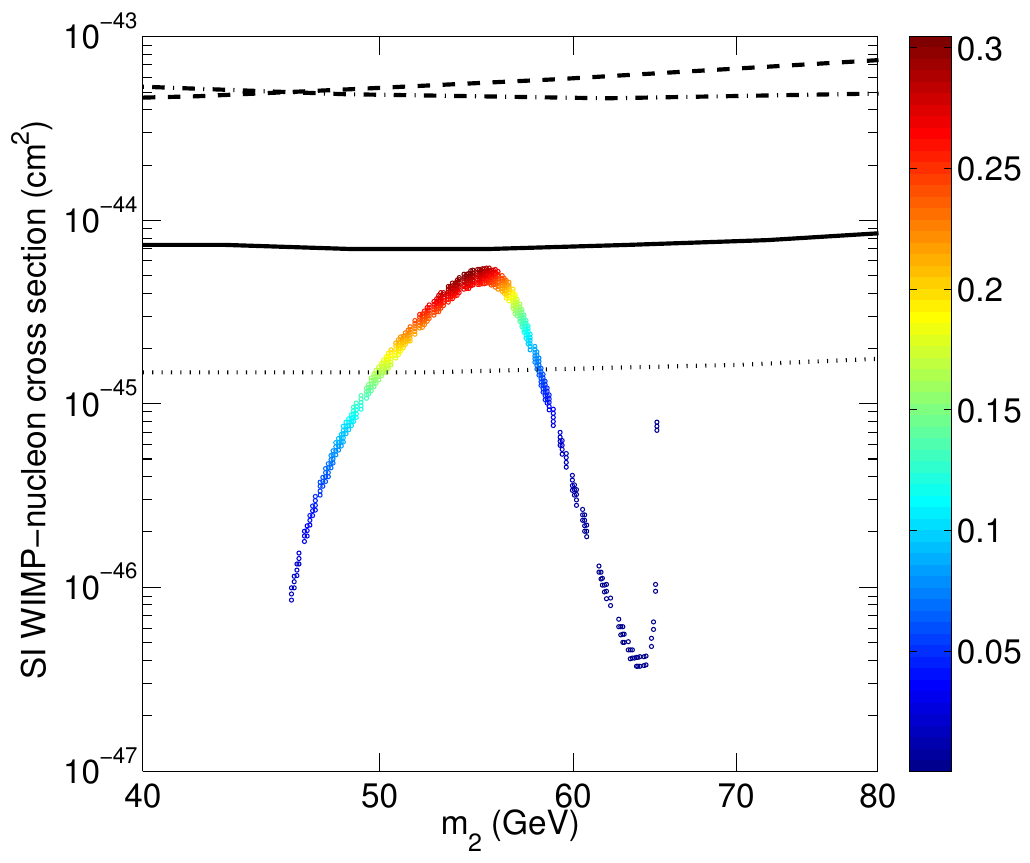}
\caption{Same as figure \ref{fig:SImh120} but for Scenario II, with $m_H=130$ GeV.}
\label{fig:SImh130S2}
\end{figure}
\begin{figure}
\includegraphics[width=0.49\textwidth]{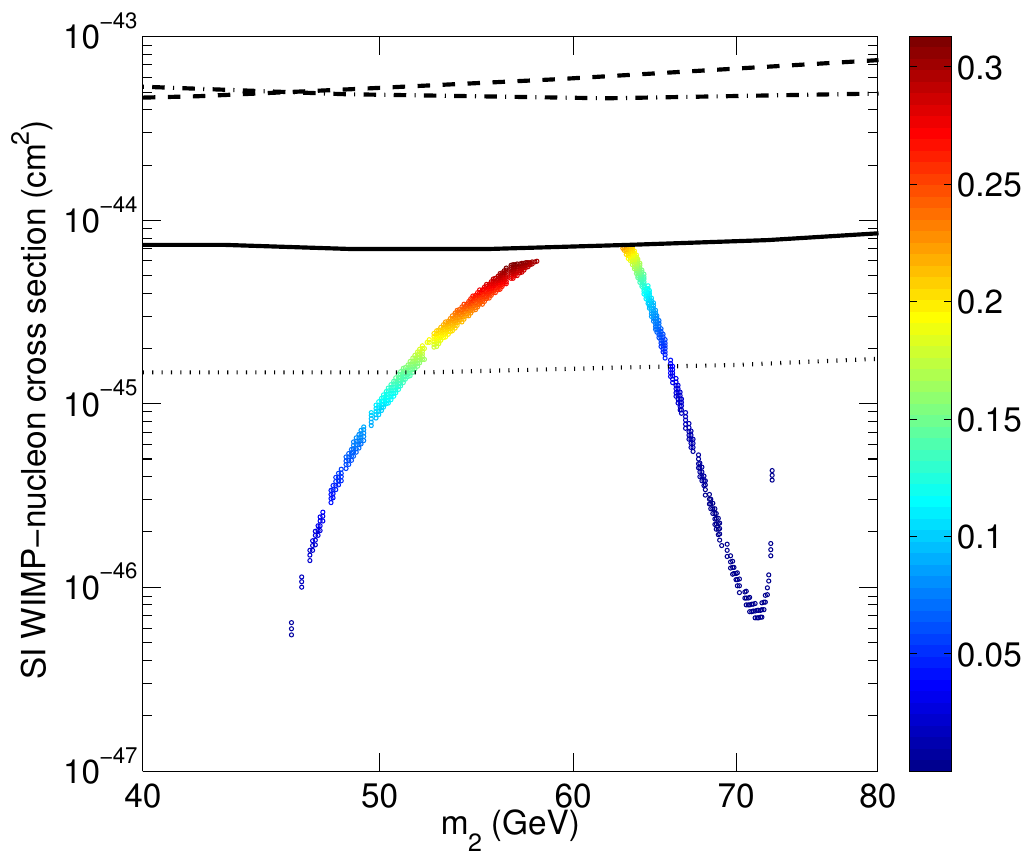}
\includegraphics[width=0.49\textwidth]{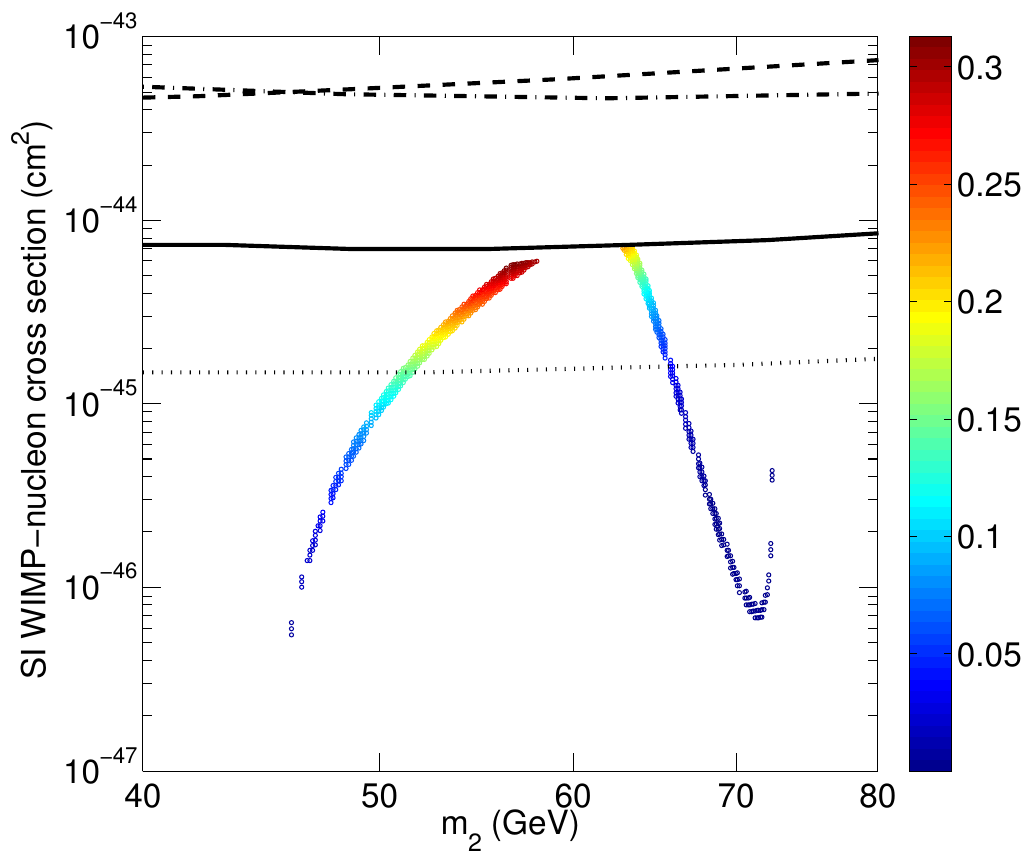}
\caption{Same as figure \ref{fig:SImh120}  but for Scenario II, with $m_H=145$ GeV.}
\label{fig:SImh145S2}
\end{figure}

%
%
 
\section{Conclusions}
\label{sec:checkout}

It is possible that a nonsequential fourth generation of leptons without quarks exists.  
We have studied the possibility that relatively light Majorana neutrinos
may be the dominant decay channel of a (composite) light Higgs boson, and that
these neutrinos, if stable, may constitute the observed dark matter
abundance of the universe. However, it seems that the area of the
parameter space where both these conditions are true is already ruled out by
direct dark matter detection experiments, the most stringent limits coming
from the spin-independent dark matter-nucleon cross section.

We investigated two possible generic realizations of the scalar sector responsible
for the mass patterns of the fourth generation leptons. In scenario I all masses 
originate from the effective SM-like Higgs field, while in scenario II, in addition to 
the SM-like Higgs field we assume that the mass of the right handed neutrino arises
from its interactions with a singlet scalar field.
In scenario I the area that produces acceptable relic density is
practically ruled out by the direct dark matter searches. Therefore this
model can not produce a realistic dark matter candidate. However, to
function as an invisible decay channel of the composite Higgs boson, the
neutrino does not need to be stable on a cosmological timescale. From
the viewpoint of collider experiments, the effect will be the same as
long as the neutrino is stable on the timescales relevant for the detector. This
could be the case even if mixing with the SM neutrinos is allowed, provided the mixing
angle is small enough. In this case, the limits from dark matter
searches obviously do not apply, and we conclude that there is a large
area of parameter space where the neutrinos are the dominant decay
channel, as shown in figures \ref{fig:mh120}-\ref{fig:mh145}.

In scenario II a large portion of the points that produce the correct
relic density lie in the area that is not ruled out by the direct dark
matter searches. Therefore the neutrino in this model, if stable, is a
promising dark matter candidate. However, in this part of the parameter
space the Higgs-neutrino coupling is weak and the invisible decay
channel will only have a subdominant effect on Higgs searches. The
composite Higgs branching ratio to the invisible channel in this case is
at most of the order of $R_\Gamma \sim 0.1$. If a light Higgs boson is
found in the 125 GeV mass range, it will be interesting to study its
partial decay widths in detail to see if there are any hints of the
presence of a subdominant invisible decay channel that would be
compatible with this particular dark matter scenario. 
Also this scenario contains areas of the parameter space where the
invisible decay channel has a large branching ratio, if we assume
that the neutrino is stable only on timescales relevant for the detector. 
Thus, if the neutrino-lifetime is suitable, the visible decays of the Higgs boson can be
highly suppressed in both models (scenarios I and II), as the invisible branching ratio is of the order of $R_\Gamma \sim 0.9$ in some parts of the parameter spaces in both models.

Depending on the value of $R_\Gamma$ the composite Higgs could still be
found in the standard search channels, if the invisible branching ratio
is not too large. Then it would just show up as having a slightly smaller than
expected production rate. On the other hand, if the invisible decay channel completely
dominates over the SM decays, the Higgs search strategy must be
altered. In this case the Higgs shows up only as missing energy in the
detector, and the most promising search channels would be monojets plus
missing energy and dileptons plus missing energy, as discussed in
\cite{Englert:2011us}. One possibility that we have not addressed in
this work is the case where the neutrino mixes with the SM neutrinos
enough to decay rapidly inside the detector. If this is the case, and
the neutrino pair is the dominant decay channel of the Higgs, then this
will obviously have significant effects on the Higgs search strategy
as well. We leave the exploration of these possibilities to future work.

\section{Appendix: Annihilation Cross Section}

Here we give the Majorana WIMP annihilation cross section to SM fermions:
\begin{eqnarray}
\sigma_{N_2N_2 \ra f \bar f}(s) =
  \frac{ G_F^2 m_W^4}{2 \pi s} \frac{\beta_f}{\beta_2} N_C^f
 && \left\{
   \frac{\sin^4\theta }{\cos^4\theta_W} |D_Z|^2 
      g_f(s,m_2, m_f ) \right.
      \nonumber \\ 
 && \left. + \frac{m_2^2 m_f^2}{m_W^4}(C_{22}^h)^2 |D_H|^2 \, s^2 \beta_2^2 \beta_f^2   
  \right\} \,,
\label{eq:fermionXsec}
\end{eqnarray}
where $\theta_W$ is the Weinberg angle, $\beta_i \equiv (1-4m_i^2/s)^{1/2}$ and %
\begin{equation}
D_X \equiv \frac{1}{s - m_X^2 + i \Gamma_X m_X}\,.
\label{DX}
\end{equation}
The factor $C_{22}^h$ accounts for the different coupling strengths of the lightest neutral particle to the SM-like Higgs as indicated in Table.~{\ref{c_table}}. Finally $N_C^\ell = 1$ for leptons and $N_C^q = 3$ for quarks and
\begin{eqnarray}
g_f(s,m_2, m_f ) \equiv && 
  \big( \sfrac{1}{3}s(s-m_f^2)\beta_2^2 + 2m_2^2m_f^2\big) 
  (v_f^2 + a_f^2)
+ \mf^2(s - 6m_2^2)(v_f^2 - a_f^2) \nonumber \\
&& -  \big( \cos^2 \theta_W 8 \frac{m_2^2 m_f^2}{m_W^2} s  
- \cos^4 \theta_W 4 \frac{m_2^2 m_f^2}{m_W^4} s^2  \big) a_f^2 
\,,
\end{eqnarray}
where $v_f = T_{3f} - 2Q_f\sin^2\theta_W$ and $a_f = T_{3f}$, where $T_3$ is the isospin and $Q$ is the charge of the fermion. The cross section in Eq.~(\ref{eq:fermionXsec}) differs from the one given in \cite{Kainulainen:2009rb} (and in \cite{Enqvist:1988dt}) as the last line in the expression of $g_f(s,m_2, m_f )$ above is absent in \cite{Kainulainen:2009rb}. Reason for this is that here we use the full $Z$-boson propagator (in the unitary gauge) when calculating the cross section, in contrast to the \cite{Kainulainen:2009rb} (and \cite{Enqvist:1988dt}) where only the part of the propagator proportional to the $g_{\mu \nu}$ has been used. This difference has only a minor impact for the DM analysis done in this work. The effect of the new terms is strongest for the heavier WIMPs $(m_2 \geq m_{top})$ for which the annihilation channel to top quarks becomes relevant. However also in that case, as we are far from the $Z$-pole, the net effect is not dramatic.

\acknowledgments
We thank T. Hapola, K. Kainulainen, C. Kouvaris, M. Nardecchia and P. Panci for insightful discussions. 



\end{document}